\documentclass[]{aa}  

\usepackage{graphicx}
\usepackage{txfonts}

\usepackage[dvipsnames]{xcolor}
\usepackage[colorlinks=true]{hyperref}

\hypersetup{
    colorlinks=true,
    linkcolor=blue,
    filecolor=blue,      
    urlcolor=blue,
    citecolor=teal
}

\newcommand{\pipes}{\textsc{bagpipes}\xspace}
\newcommand{\Mstar}{\ensuremath{\mathrm{M}_\star}\xspace}

\newcommand{\sig}{\ensuremath{\sigma}\xspace}

\newcommand{\eazy}{\textsc{eazy}}

\newcommand{\tng}{\textsc{Illustris-TNG100}\xspace}
\newcommand{\eagle}{\textsc{Eagle}\xspace}
\newcommand{\flamingo}{\textsc{Flamingo}\xspace}

\newcommand{\magneticum}{\textsc{Magneticum}\xspace}

\newcommand{\simba}{\textsc{Simba}\xspace}
\newcommand{\shark}{\textsc{Shark}\xspace}
\newcommand{\galform}{\textsc{Galform}\xspace}

\newcommand{\Mgten}{\ensuremath{\mathrm{\Mstar>\rm 10^{10}M}_\odot}\xspace}

\newcommand{\Mgnine}{\ensuremath{\mathrm{\Mstar>\rm 10^{9.5}M}_\odot}\xspace}

\begin{document}

   \title{Exploring over 700 massive quiescent galaxies at z = 2–7: Demographics and stellar mass functions}
   \titlerunning{high-z quiescent galaxies}

   \author{William M. Baker
          \inst{1}\fnmsep\thanks{william.baker@nbi.ku.dk}
          \and
          Francesco Valentino \inst{2,3}
          \and 
          Claudia del P. Lagos \inst{4,5, 2}
          \and
          Kei Ito \inst{2,3}
          \and
          Christian Kragh Jespersen \inst{6}
          \and
          Rashmi Gottumukkala\inst{2,7}
          \and
          Jens Hjorth \inst{1}
          \and
          Danial Langeroodi \inst{1}
          \and
          Aidan Sedgewick \inst{1}
          }

   \institute{DARK, Niels Bohr Institute, University of Copenhagen, Jagtvej 155A, DK-2200 Copenhagen, Denmark 
        \and
        Cosmic Dawn Center (DAWN), Copenhagen, Denmark
        \and 
        DTU Space, Technical University of Denmark, Elektrovej 327, DK-
2800 Kgs. Lyngby, Denmark
        \and
        International Centre for Radio Astronomy Research (ICRAR), M468, University of Western Australia, 35 Stirling Hwy, Crawley, WA 6009, Australia
        \and
        ARC Centre of Excellence for All Sky Astrophysics in 3 Dimensions (ASTRO 3D)
        \and
        Department of Astrophysical Sciences, Princeton University, Princeton, NJ 08544, USA
\and
Niels Bohr Institute, University of Copenhagen, Jagtvej 128, 2200 Copenhagen N, Denmark
        }


 
  \abstract
   {High-redshift ($z>2$) massive quiescent galaxies are crucial tests of early galaxy formation and evolutionary mechanisms through their cosmic number densities and stellar mass functions (SMFs). We explore a sample of 743 massive ($\rm M_*> 10^{9.5}M_\odot$) quiescent galaxies from $z=2-7$ in over 800 arcmin$^2$ of NIRCam imaging from a compilation of public JWST fields (with a total area $>$ 5 $\times$ previous JWST studies).
   We compute and report their cosmic number densities, stellar mass functions, and cosmic stellar mass density. We confirm a significant overabundance of massive quiescent galaxies relative to a range of cosmological hydrodynamical simulations and semi-analytic models (SAMs). 
   We find that no simulations or SAMs accurately reproduce the SMF for massive quiescent galaxies at any redshift within the interval $z=2-5$. This shows that none of these models' feedback prescriptions are fully capturing high-z galaxy quenching, challenging the standard formation scenarios. We find a greater abundance of lower-mass ($\rm M_*<10^{10}M_\odot$) quiescent galaxies than previously found, highlighting the importance of sSFR cuts rather than simple colour selection. We show the importance of this selection bias, alongside individual field-to-field variations caused by cosmic variance, in varying the observed quiescent galaxy SMF, especially at higher-z. We also find a steeper increase in the cosmic stellar mass density for massive quiescent galaxies than has been seen previously, {with $\rho_*\propto (1+z)^{-7.2\pm0.3}$}, indicating the dramatic increase in the importance of galaxy quenching within these epochs. 
   }


   \maketitle
%

\section{Introduction}
Massive quiescent galaxies are crucial tests of our understanding of early galaxy formation and evolutionary mechanisms. Simply "counting" their number within a given volume (i.e. computing their cosmic number densities) has strong implications as to the strength and effect of feedback processes on galaxy evolution \citep[e.g.][]{Somerville2015,Beckmann2017,Merlin2019,Donnari2021,Donnari2021a,Baker2025a, DeLucia2024,Lagos2025}. 
These massive quiescent galaxies are the descendents of earlier star-forming galaxies that have undergone the process of quenching, after which galaxies cease forming new stars \citep[at least in any significant number; for a review see][]{Man2018,Curtis-Lake2023, DeLucia2025}. 

Many possible quenching mechanisms have been proposed such as stellar feedback \citep{Larson1974,Dekel1986, Veilleux2005}, AGN feedback \citep{Silk1998,Croton2006, Fabian2012}, starvation \citep{Peng2015, Trussler2020, Baker2024}, morphological {quenching} \citep{Martig2009}, cosmic rays \citep{Ruszkowski2023}, environmental {quenching} \citep{Alberts2024}, and many more \citep{Man2018}. It can be helpful to split quenching mechanisms based on the stellar masses of the galaxies in question {as it is believed that different mechanisms are responsible for various galaxy subpopulations}. This becomes even more important when investigating high-redshift galaxy quenching as it necessarily places more constraints on both the masses of galaxies and the timescales for quenching because of the younger age of the Universe at these redshifts.

At least more locally, galaxy quenching has been shown to be roughly split into two categories based on stellar mass and environment \citep{Peng2010}, mostly corresponding to internal versus external quenching mechanisms \citep[also seen in simulations, e.g.][]{Lim2025}. The external mechanisms (under the broad name of environmental quenching) correspond to the region in which the galaxy is present, with more overdense regions having a greater proportion of passive galaxies found \citep[][]{Dressler1980}. This is traditionally thought to affect lower-mass galaxies ($M_\star<10^{10}\ \rm{M_\odot}$) more than the high-mass galaxies ($M_\star>10^{10}\ \rm{M_\odot}$) due to their shallower gravitational potentials, making them less able to hold on to their gas reservoirs, whilst dealing with processes such as ram-pressure stripping or harassment \citep{Gunn1972,Peng2010}. 
For more massive galaxies, generally internal quenching mechanisms are thought to be the most important, especially in cosmological simulations \citep[see for a review see][but mergers and environment can still play a role in certain cases]{Somerville2015,Man2018, Vogelsberger2020}.

{Prior to JWST, spectroscopically confirmed massive quiescent galaxies were limited to a handful of studies with the redshift frontiers being continually pushed  higher \citep[e.g.][]{Dunlop1996, Cimatti2004, Daddi2005, Glazebrook2017,Belli2019, Valentino2020}}
However, since the advent of the JWST we have been finding many more high-z massive quiescent galaxies than previously found, and to even higher redshifts \citep[e.g.][]{Carnall2024, Nanayakkara2025, Nanayakkara2024, Baker2025a, deGraaff2025, Weibel2024qgal, Wu2025}. This has resulted in high number densities that disagree with many aspects of theory \citep{Schreiber2018,Merlin2019,Carnall2023, Valentino2023, Baker2025a}. This was shown in some of the first data taken with JWST \citep[e.g.][]{Carnall2023, Long2024, Russell2024, Alberts2024}, but, due to the small regions probed, cosmic variance was shown to be an increasingly important factor \citep{Steinhardt2021, Valentino2023}. Compared with theory, it is also important to take into account the effects of cosmic variance within cosmological simulations, as in \citet{Baker2025a}, where they found that although this affects number densities, there is still a 3$\sigma$ disagreement at z>3. The observational overabundance remains even when using spectroscopy to better ensure the fidelity of the quiescent galaxy sample \citep{Baker2025a}.
This has confirmed that we are observing more of these high-z quiescent galaxies than we find in our best cosmological simulations \citep{Valentino2023,Lagos2025, Baker2025a}, i.e. more massive quiescent galaxies exist at high-z than are predicted by our current models and theory.

If we classify high-z massive quiescent galaxies as those with stellar masses above $\rm M_\star>10^{9.5}M_\odot$\footnote{
We note that the mass ranges probed are also important; the JWST has uncovered a population of low-mass ($ M_\star<10^{9.0}M_\odot$) "mini" or "rapidly" quenched galaxies at z>5 \citep{Strait2023, Looser2024, Baker2025b}. These galaxies are not standard quiescent galaxies because while they have visible Balmer breaks they also have low masses and blue colours, an indication of very recent quenching $\rm t_{quench}\leq30Myr$.}, then when
comparing to cosmological hydrodynamical simulations and semi-analytic models (SAMs), another key aspect to probe is not only the number densities, but also the number of quiescent galaxies of a given stellar mass within a given area. This gives the well-studied stellar mass function \citep[SMF, e.g.][]{Grazian2015,Davidzon2017,Leja2020, Weaver2023, Weibel2024}, only for high-z quiescent galaxies. Stellar mass functions are crucial for understanding many aspects of galaxy evolution, particularly the separation into separate star-forming and quiescent galaxy subpopulations. However, high-z exploration of the SMFs has previously been conducted based solely on colour selection \citep[e.g.][]{Weaver2023}, rather than full Bayesian spectral energy distribution (SED) modelling. As has previously been shown in multiple works from both observations and simulations \citep[][]{Antwi-Danso2023,Lovell2023,Baker2025a}, this is likely to miss a significant number of quiescent galaxies at high-z. 

Generally cosmological hydrodynamical simulations and SAMs are designed to reproduce the stellar mass function at low redshifts, but at high-z predictions deviate substantially for massive quiescent galaxies \citep{Lagos2025} compared to previous observations \citep[][]{Weaver2023}. The different predictions for high-z number densities and stellar mass functions from various cosmological simulations and SAMs were explored in depth in \citet{Lagos2025}. They showed that there were significant differences between the observations and simulations, as well as significant differences between the simulations and the SAMs themselves. This opens up an exciting new avenue for testing our understanding of galaxy evolutionary physics using the abundances and masses of high-z quiescent galaxies.

However, on the observational side, we have yet to fully exploit the wide area of publicly available JWST imaging, the largest area explored so far being 145 arcmin$^2$ in \citet{Valentino2023}. 
Currently we have one spectroscopically confirmed quiescent galaxy above $z=5$ \citep{Weibel2024qgal, Valentino2025} and a handful above $z=4$ \citep{Tanaka2019,Carnall2023Nature,Carnall2024,Kakimoto2024,UrbanoStawinski2024,deGraaff2025, Baker2025a, Barrufet2025,Wu2025, Nanayakkara2025, Antwi-Danso2025}. To properly compute number densities, we require a comprehensive understanding of the area probed, which is complex in the case of spectroscopy due to target selection. For this we require a photometric approach which also brings the benefit of greatly enhanced sample sizes, albeit without the exquisite detail of a spectrum.

In this paper, we will explore a comprehensive photometric sample of $z>2$ quiescent galaxies. We will compute their cosmological number densities, stellar mass functions, and report high-z candidates. Our combined area consists of 816 arcmin$^2$ of publicly available JWST imaging, which is more than 5$\times$ larger than \citet{Valentino2023}.

\section{Data}

As we want to compute number densities and stellar mass functions, it is crucial that we probe a large area of JWST imaging. To do this, we use a combination of publicly available JWST fields. These correspond to the Cosmic Assembly Near-infrared Deep Extragalactic Legacy Survey \citep[CANDELS,][]{Grogin2011, Koekemoer2011} fields. We use the Cosmic Epochs Early Release Survey \citep[CEERS, DD-ERS 1345][]{Finkelstein2025}, Public Release IMaging for Extragalactic Research \citep[PRIMER, GO 1837][]{Dunlop2021, Donnan2024}, and the JWST Advanced Deep Extragalactic Survey \citep[JADES, GTO 1180, 1181, 1210, 1287, 1215][]{Eisenstein2023, Rieke2023}. This consists of multiband imaging in EGS, UDS, COSMOS, GOODS-S and GOODS-N \citep[][]{Giavalisco2004, Grogin2011, Koekemoer2011}. The idea behind this is that we are more robust to cosmic variance using a combination of different fields from different regions of the sky. This culminates in a combined total area of $\rm 816.9\; arcmin^2$.

We use the photometric data from the DAWN JWST Archive \citep[DJA][]{Valentino2023, Heintz2025}\footnote{https://dawn-cph.github.io/dja/index.html} a publicly available repository of reduced JWST imaging and spectroscopy. The photometric data was processed with the \textsc{GRIZLI} \citep{Brammer2023} code. See \citet{Valentino2023} for more details about the data reduction and photometric extraction process.

For our colour selection with \eazy\ \citep{Brammer2008} we use all available photometric data including HST, NIRCam and MIRI data. For SED modelling with \pipes \citep{Carnall2018} we use all available JWST data (both NIRCam and, where available, MIRI). We use the corrected 2nd aperture flux values, corresponding to an aperture of 0.7\arcsec~in diameter. These follow the standard DJA format \citep[see][for details]{Valentino2023} hence are not PSF-matched, but have been shown to provide accurate stellar population properties for high-z quiescent galaxies in \citet{Ito2025b}.
In this work, we use a \citet{Kroupa2001} IMF \footnote{We note that the simulations use a \citet{Chabrier2003} IMF, but that the difference between the two is minor and less than the errors in obtaining stellar masses and SFRs from photometric SED modelling.} and \citet{PlanckCollaboration2020} cosmology throughout.

\section{Methods and sample selection}
Our selection criteria consists of three steps. Step 1 is an expanded UVJ colour selection via the use of the template fitting code \eazy\ \citep{Brammer2008}. Step 2 is a specific star-formation rate (sSFR) cut after more detailed SED modelling with \pipes \citep{Carnall2018, Carnall2019}. Finally, step 3 is visual inspection of the SED fits in order to remove poorly fit galaxies or clear contaminants. In addition, for $z>5$ targets we also implement a brown dwarf removal cut based upon colour-colour selection \citep{Langeroodi2023, Kokorev2024}. 

\subsection{Template fitting with \eazy}

We start by using the template fitting code \eazy\ \citep{Brammer2008} to compute redshifts alongside rest-frame UVJ colours for all the sources within our fields. We use the standard templates included in the DJA analysis pipeline \citep{Valentino2023}, alongside the standard eazy parameters. 
\eazy\ does template fitting across a redshift grid by finding the best-fit linear combination of templates at each redshift, then from the $\chi^2s$ of the best-fit SEDs, it finds the best-fit redshift. It then extracts rest-frame colours and masses from the best-fit SEDs.
We select galaxies with a photometric redshift $z>2$ and with a stellar mass \Mstar$>10^{9.3}M_\odot$. We adopt this lower mass cut to account for EAZY potentially underestimating stellar masses.

{For our initial colour selection, we use the criteria of \citet{Baker2025b}. This selection recovers many high-redshift quiescent galaxies that fall outside a standard colour selection box \citep[such as][]{Schreiber2018}. Our selection corresponds to the UVJ colours}
\begin{equation}
    U-V\ > \ 1.30 \ (V-J) - 0.20.
\end{equation}

This has been shown to fully recover 18 massive high-z spectroscopically confirmed quiescent galaxies \citep{Baker2025a}. {We reiterate that our key} cut will be based on sSFR which will remove any extra contaminants introduced by this more relaxed approach. 
This cut reduces our sample to 15,560 galaxies.

\subsection{SED modelling with \pipes}

To model our candidate quiescent galaxies we employ the Bayesian SED modelling code \pipes \citep{Carnall2018, Carnall2019}. See Appendix \ref{app.s.pipes_setup} for the exact details of our setup.

To select quiescent galaxies from those fit we utilise a traditional cut based on sSFR \citep[e.g.][]{Franx2008,Gallazzi2014,Schreiber2018,Carnall2023} which corresponds to 
\begin{equation}
\label{eq:ssfr_cut}
   \rm  sSFR\ \leq 0.2/t_{age}
\end{equation}
where $\rm t_{age}$ is the age of the universe at the observed redshift.
In addition, as we are after massive quiescent galaxies, we select those with \Mstar $>10^{9.5}M_\odot$ (except for our analysis of the SMFs in Sec. \ref{s.smfs}).
This cut reduces our sample from 15,560 galaxies to 882 galaxies.

\subsection{Visual inspection and Brown Dwarf cut}
For candidate $z>5$ quiescent galaxies with JWST, brown dwarfs (BD) become interlopers \citep[e.g.][]{Kauffmann2020}. To deal with this problem, we use the selection criteria of \citet{Langeroodi2023} based on BD templates. This corresponds to a cut of F200W$-$F150W>0.20. We combine this with a cut from \citet{Kokorev2024} which was based on selecting for the so called 'Little Red Dots' \citep[LRDs, e.g.][and many more works]{Matthee2024}. There is still much debate on the nature of LRDs as to whether they are AGN \citep[e.g.][]{Inayoshi2025} or quiescent galaxies \citep[e.g.][]{Baggen2024}. 

In light of this, we settle for removing galaxies with clear V-shaped SEDs via visual inspection as possible LRD or BD contaminants. Most high-z quiescent galaxies \citep[e.g. those seen in ][]{Weibel2024,deGraaff2025,Baker2025a, Nanayakkara2025} show blue slopes redwards of the Balmer break so the removal of V shaped SEDs should not affect our selection of these types of massive quiescent galaxies. 
We also remove candidate quiescent galaxies with insufficient filter coverage to properly probe the Balmer break, as we cannot be certain of those galaxies quiescent nature (this is however, a negligible number of galaxies)
\footnote{As an example, we remove galaxies with only F200W, F277W, F356W and F444W, if the break is blueward of F200W. We also remove any galaxy with fewer than 4 filters (there are a handful of well-constrained $z=2$ galaxies with 4 filters that properly probe the full NIRCam range). We note that this only applies to the \pipes fitting - we use the full JWST + Hubble filters in the initial \eazy \ based selection.}.
These cuts reduce our sample from 882 galaxies to 743 galaxies.

\begin{figure}
    \centering
    \includegraphics[width=0.9\linewidth]{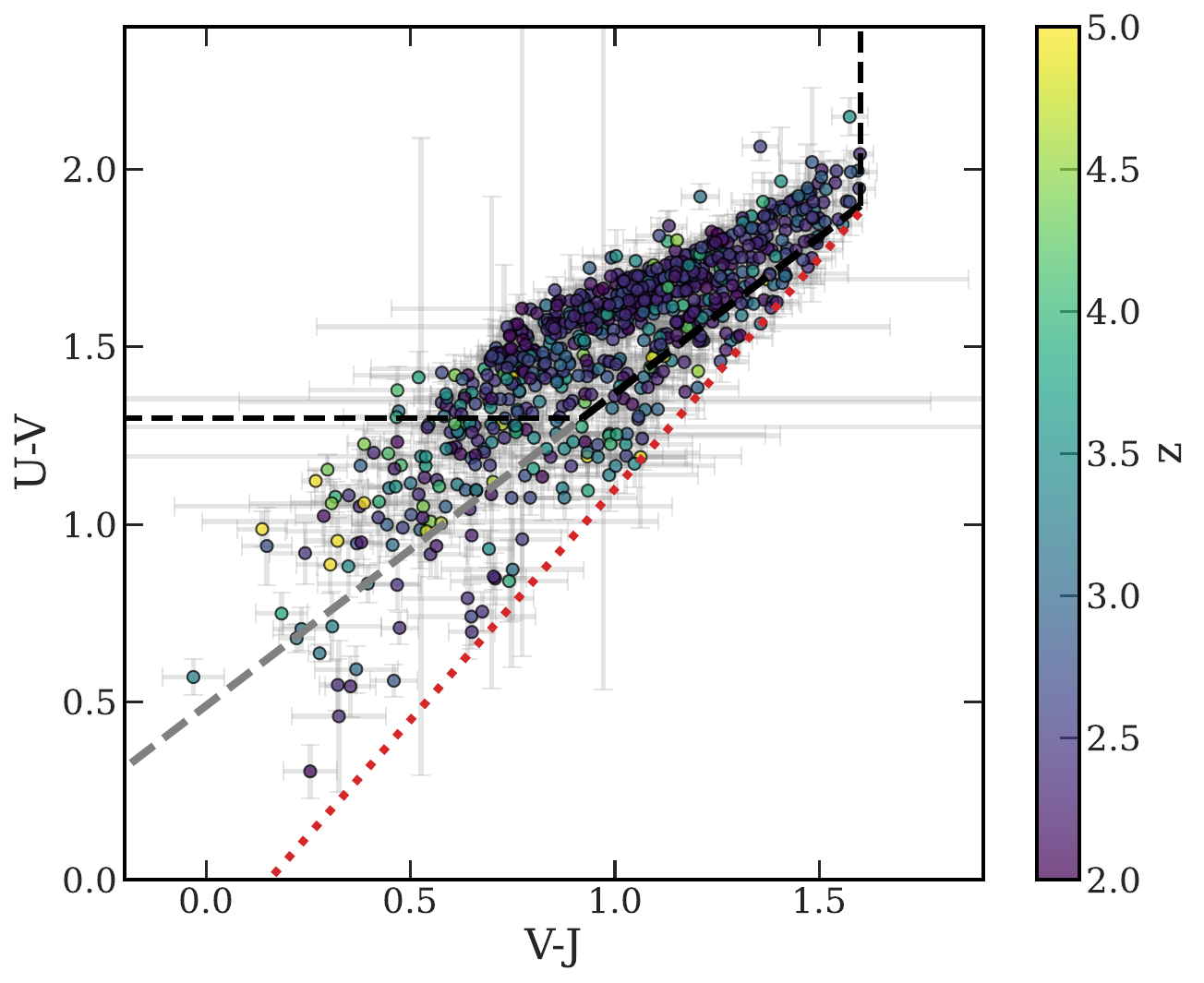}
    \caption{Upper panel: U-V and V-J colour selection colour-coded by redshift for the massive quiescent galaxies in our full sample. The black line corresponds to the \cite{Schreiber2015} quenching criterion with the grey line addition of the \cite{Belli2019} fast quenching criterion. The red dotted line corresponds to the selection criteria from \cite{Baker2025a} which we use as our initial selection criteria in this work. 
    }
    \label{fig:uvj}
\end{figure}

Our final full sample consists of 743 massive quiescent galaxies with $M_\star > 10^{9.5}M_\odot$ and with \eazy\ photometric redshifts of $z>2$. 
Fig. \ref{fig:uvj} shows the distribution of our final sample in UVJ colour space, colour-coded by redshift. We can see that our quiescent galaxy sample primarily occupies the traditional criteria, but with significant numbers within the new \citet{Baker2025a} criteria. These would have been missed by strict UVJ cuts, but have been selected by our new colour selection with confirmation from the sSFR cut. We explore F356W magnitude versus stellar mass binned in redshift for the full sample in Appendix \ref{app.s.completeness}.

\section{Number densities}

\begin{figure}
    \centering
    \includegraphics[width=1\linewidth]{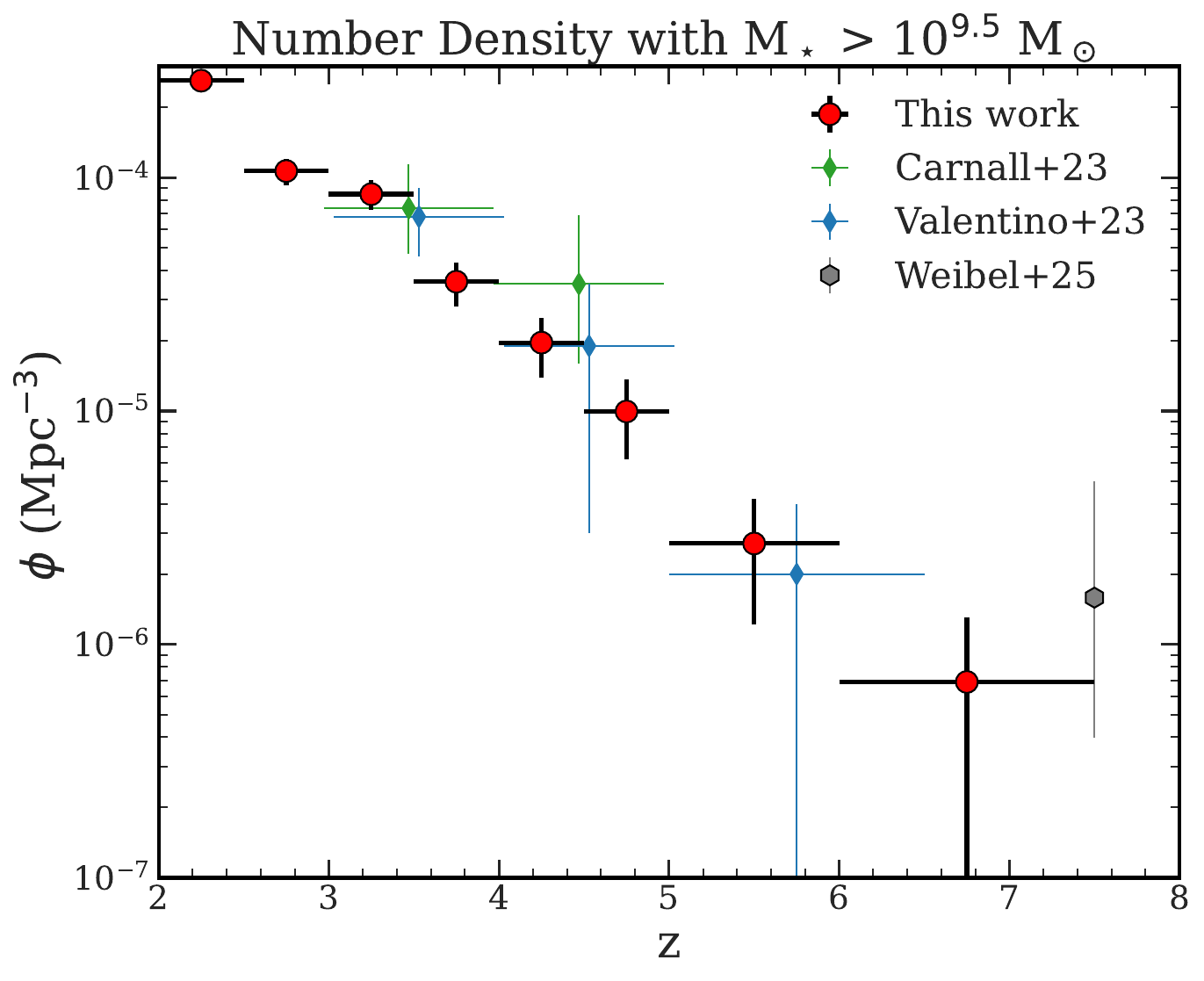}
    \caption{Quiescent galaxy number densities of those with \Mstar>$\rm 10^{9.5}M_\odot$ from our sample (red points). We compare to \citep{Carnall2023} green points, \citet{Valentino2023} blue points, and the single galaxy from \citet{Weibel2024qgal} as the grey point. {We offset the points ever so slightly in redshift for the \citet{Valentino2023} and \citet{Carnall2023} number densities so as to make their errors visible within the figure.}}
    \label{fig:number-density}
\end{figure}

The first aspect to explore for the full sample of massive quiescent galaxies is their number densities. In Fig. \ref{fig:number-density} we show the massive (\Mstar>$\rm 10^{9.5}M_\odot$) quiescent galaxy number density versus redshift for our study (red markers) and a compilation of other equivalent studies from the literature. 
The number density is computed as the number of massive quiescent galaxies within a given redshift range divided by the total co-moving volume within that redshift range.
The area of each field is computed using the total area of the F444W mosaic for each field, as this is used in computing the detection image in the DJA pipeline \citep{Valentino2023}. 
The volume is then computed assuming 
\begin{equation}
\rm V=\frac{\Omega}{3}\times (d_1^3-d_2^3).
\end{equation}
For more details on this method see \citet{Baker2025a} where it is described in detail.
As can be seen in Appendix Fig. \ref{fig:completeness}, even our faintest massive quiescent galaxy is significantly brighter than our shallowest survey limiting magnitude \footnote{Which following \citet{Donnan2024} we take the 5$\sigma$ global depth in F444W to be 27.9 mag in the shallowest region of the Primer-UDS field} so we do not have to worry about Malmquist bias due to our mass cut of \Mstar>$\rm 10^{9.5}M_\odot$. 

Our errors for the number densities are a combination of three different causes. First we bootstrap the number density calculation for variations in the \eazy\ redshifts. This enables galaxies to shift redshift bin. This has an almost negligible effect on our number densities. Secondly, we incorporate the effects of cosmic variance. We follow the prescription of \citet{Jespersen2025}, calibrating the cosmic variance based on a combination of the cosmic variance calculator by \citep{moster2011} and added constraints from the \texttt{UniverseMachine} simulations \citep{Behroozi2019} at the highest redshifts. This directly incorporates realistic effects of scatter due to assembly bias \citep{Jespersen2022, Chuang2024}.
This has strong effects on individual fields, but it is less of an issue here due to the combination of separate sightlines and our (comparatively) large area. Thirdly, we include Poisson counting error which becomes the main source of error in the high-z bins. All of these errors are combined in quadrature to produce our overall errors. We follow the same approach when computing errors throughout this work.

The results of this is shown in Fig. \ref{fig:number-density}.
We see that we obtain a rapid decline with redshift as would initially be expected.
We limit the amount of comparisons here as we wish to (at least) match the exact mass cut used in the literature (see Fig. \ref{fig:number-density-10} for many more comparisons to observations and simulations). 
The most comparable observational study is that of \citet{Valentino2023} as they also select massive quiescent galaxies with \Mstar>$\rm 10^{9.5}M_\odot$ enabling a more like-for-like comparison. 
\citet{Valentino2023} had a field size of 145 arcmin$^2$, whereas in this work we have an area of 816 arcmin$^2$ which is approximately 5$\times$ the size. This gives us significantly greater number statistics.
They used UVJ selection, but included everything within 1 \sig, in effect expanding the criteria. 

Our number densities remain mostly consistent with this work despite our much larger field size and redshift range containing $\sim$ 9$\times$ as many quiescent galaxies as in that work. This shows the benefit of selecting multiple different fields in trying to deal with the effects of cosmic variance. We also show the results of \citet{Carnall2023}, which shows an excess in the higher redshift bin due to an overdensity of massive quiescent galaxies in CEERS \citep{Jin2024,Valentino2023}. 

We also compare the number density obtained in \citet{Weibel2024qgal}. This is based on a single $z=7.3$ massive quiescent galaxy obtained as part of the RUBIES program \citep{deGraaff2024} \footnote{We note that this galaxy has been recovered as part of our photometric sample in this work.}. A key difficulty for that work is that they found a single galaxy and it is not entirely apparent what volume should be considered for it.
We find a number density slightly lower than was found in \citet{Weibel2024qgal} but it remains consistent with our \Mgnine result.

\subsection{Number densities above \Mstar>$\rm 10^{10}M_\odot$}
\label{s.number_densities10}
\begin{figure}
    \centering
    \includegraphics[width=0.9\linewidth]{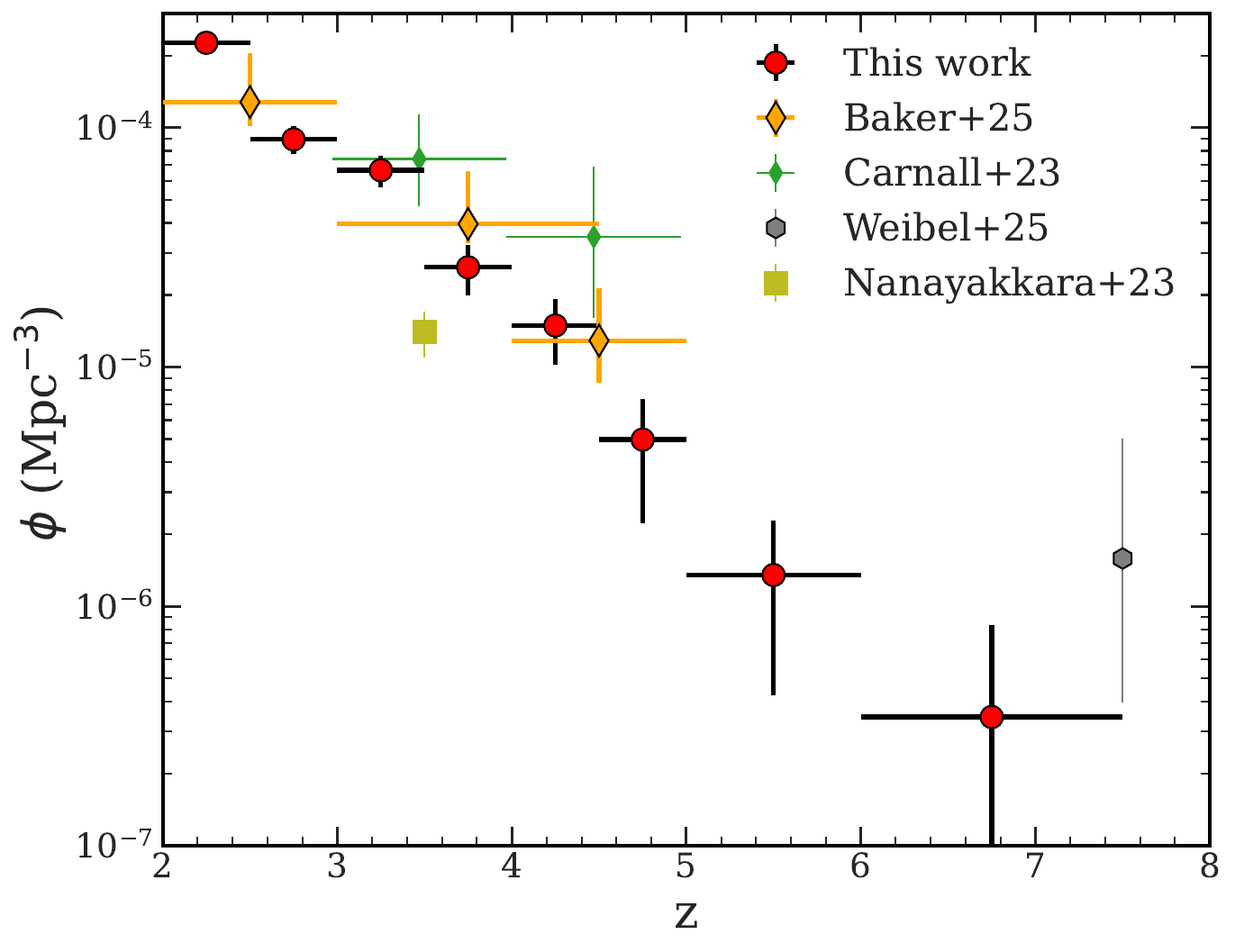}
    \includegraphics[width=0.90\linewidth]{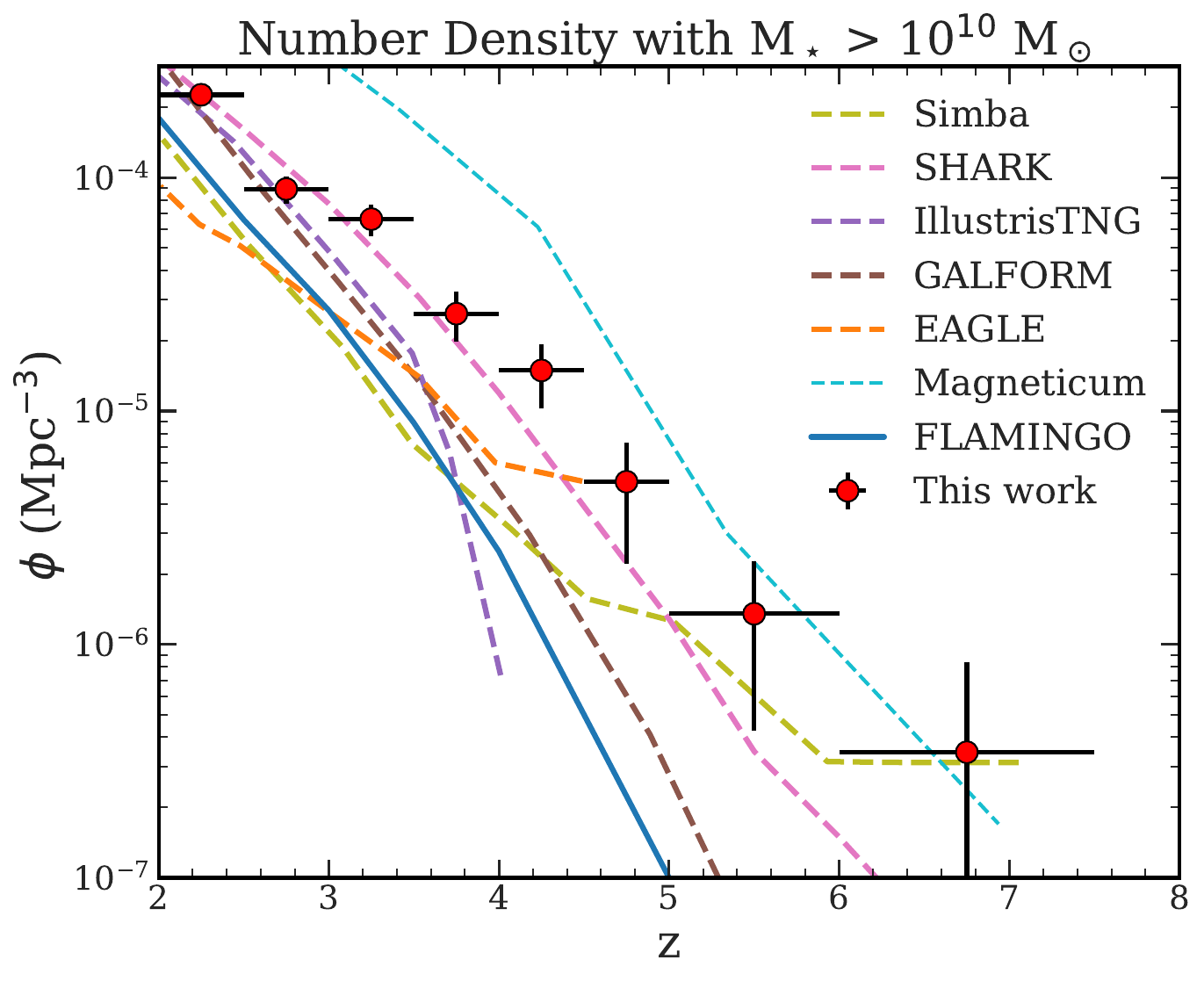}
    \includegraphics[width=0.90\linewidth]{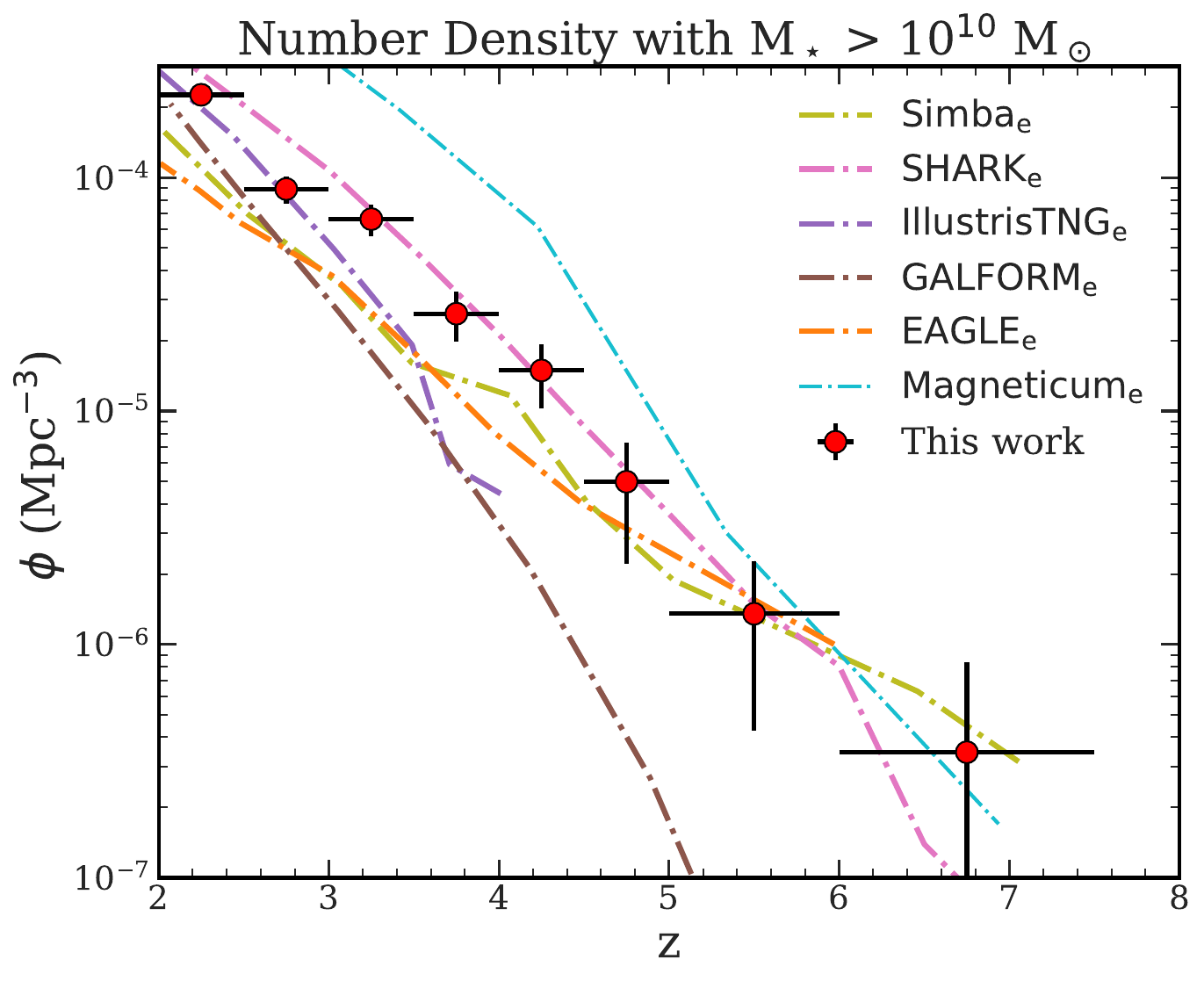}
    \caption{Quiescent galaxy number densities for \Mstar>$\rm 10^{10}M_\odot$ from our sample (red points). 
    Upper panel: a comparison to other observational studies,  \citet{Carnall2023}, green points, a spectroscopically corrected sample, \citet{Baker2025a}, orange points, a spectroscopic sample, \citet{Nanayakkara2025}, light green point, and the \citet{Weibel2024qgal} single galaxy point. Middle panel: a comparison to 5 cosmological hydrodynamical simulations, \eagle, \tng, \simba, \flamingo, and \magneticum \citep[][]{Lagos2025,Baker2025a, Remus2025} and 2 SAMs, \shark and \galform. Bottom panel: 4 cosmological hydrodynamical simulations and 2 SAMs, but with Gaussian scatter included in the stellar masses and star-formation rates. We see that simulations struggle to reproduce the observed number densities at most redshifts.}
    \label{fig:number-density-10}
\end{figure}

Due to the number of simulations with limited resolution, and because at lower redshift it is often used as a benchmark, it is also useful to explore the number density of massive quiescent galaxies with \Mstar>$\rm 10^{10}M_\odot$. This naturally reduces our sample size to 614 massive quiescent galaxies.
We explore their number densities in Fig. \ref{fig:number-density-10}. We compare them to a selection of other comparable observations and simulations. 
 
Fig. \ref{fig:number-density-10}, upper panel, shows our observed number density versus several comparable studies. 
Observationally, due to the large volume area probed by our study, we are able to use more finely grained binning enabling a greater understanding of the evolution of the number density. Comparing with the literature, there are few JWST-era studies on number densities in fields with large volumes, and most are based on single fields that are likely to be significantly affected by cosmic variance \citep[e.g.][]{Steinhardt2021, Carnall2023, Long2024, Valentino2023, Jespersen2025}.
In spite of this, we find good agreement with the results of \citet{Baker2025a}, which was based on a small survey area of GOODS-S (also included as a field within this work) but with a correction to the photometric number densities based on their spectroscopic sample. {They used a traditional UVJ selected photometric sample \citep[selected via][]{Ji2024} as the baseline, but rescaled the quiescent galaxy number densities based on the confirmation/accuracy rate when compared to a spectroscopic sample, which itself was selected primarily by an evolving sSFR cut.}
We find significantly more massive quiescent galaxies than those found by \citet{Nanayakkara2025}, but their result was based on spectroscopy in a smaller area. Again we see the effect of the overdensity in CEERS on the \citet{Carnall2023} number densities which are significantly higher than ours at $z=4-5$ (but based on only a handful of galaxies in a small area).
As shown in the previous section, we again find a number density lower than was found in \citet{Weibel2024qgal} for our \Mgten number densities.

One of the key benefits of the \Mgten selection is we can more cleanly compare with simulations and SAMs, as most predictions presented in the literature adopt that mass threshold.
We first compare to the \flamingo simulations \citep{Schaye2023, Kugel2023} using the number densities computed in \citet{Baker2025a}. {These \flamingo number densities from \citet{Baker2025a} were computed from the higher resolution 1Gpc$^3$ \flamingo simulation box based on the same evolving sSFR criteria adopted in this work (e.g. Eq. \ref{eq:ssfr_cut}).} This enabled \citet{Baker2025a} to calculate the effect of cosmic variance on the simulations to up to 3\sig. We do not include the errors from that work here, as they were computed to match the observed area within that work. However, we do include the number density from the simulation as the blue line in the middle panel of Fig. \ref{fig:number-density-10}. We see that we obtain significantly higher number densities than are found in \flamingo at all redshifts probed, which is consistent with the findings in \citet{Baker2025a}.

The next stage is to compare with other cosmological simulations with different setups, feedback prescriptions, box-sizes, and sub-grid physics. We use the results of \citet{Lagos2025} which consists of 3 cosmological hydrodynamical simulations and 2 SAMs\footnote{We restrict the analysis here to the simulations used in \citet{Lagos2025} which make their data publicly available}. The cosmological simulations are \eagle \citep{Schaye2015, Crain2015}, \tng  \citep{Pillepich2018, Springel2018}, and \simba \citep{Dave2019}. The SAMs consist of \shark \citep{Lagos2018, Lagos2024} and \galform \citep{Lacey2016}. For full details on the simulations, see \citet{Lagos2025}. Here, we modified the default sSFR selection in \citet{Lagos2025}-- based on fixed sSFR thresholds-- to match our evolving time-dependent selection of Eq.~\ref{eq:ssfr_cut}.

The comparison is shown in the middle panel of Fig. \ref{fig:number-density-10}, where the simulations are denoted by dashed lines. Upper limits from the simulations are not plotted, hence why some tracks end abruptly (these can be envisaged as horizontal lines from the end of each track).

As shown in \citet{Lagos2025}, \tng  \citep{Pillepich2018, Springel2018} shows a significant lack of massive quiescent galaxies above z=4 (in fact there are none in the $(100\,\rm Mpc)^3$ volume, hence it becomes an upper limit of around $\sim10^{-6}$), which totally disagrees with the observations. However, note that \tng does manage to reproduce the number density of massive quiescent galaxies at $z=2-3$ with \Mgten.
{\eagle fails to match any of our observed number densities at every redshift whilst \simba also fails at almost all redshifts.
The same is found for SAM \galform, although it does reproduce the number densities at $z=2-3$ (like \tng).}

On the other hand, the SAM \shark \citep{Lagos2018, Lagos2024} appears to agree better with our observed number densities. It remains mostly consistent with our observations up to $z\sim$6 and only substantially starts to deviate at $z\geq$6 which is the most unconstrained with simulations. This is likely a result of the AGN feedback prescriptions within \shark, consisting of a radiative and jet mode feedback \citep[which can be triggered by mergers or disc instabilities, enabling more possible triggers,][]{Lagos2025}. 

{In addition to the aforementioned simulations, recent studies \citep[e.g.][]{Remus2025, Kimmig2025, Chittenden2025} have shown that the cosmological hydrodynamical simulation \magneticum appears to be able to produce many more high-z quenched galaxies than other simulations. Due to this, we also include a matched comparison in this work, whereby the \magneticum number densities are given by the light blue/turquoise lines in Fig. \ref{fig:number-density-10}.  We see that whilst it is able to produce enough quiescent galaxies at $z=5-7$, it significantly overproduces them at lower redshifts (e.g. $z=2-5$), where our constraints are most stringent. This is likely due to the different black hole feeding growth caused by the simulation treating hot and code accretion modes separately, with a boost to the cold mode by around a factor of 100 \citep{Gaspari2013}. This ensures high accretion onto the BH at early times, helping boost quenching above $z=5-7$, but likely results in there being so much cold gas at $z=2-5$ that the BH is still fed very efficiently, driving overly strong quenching at these lower redshifts which is in significant disagreement with our observations.  }

The next stage is to try and introduce realistic observational errors in the simulations to attempt to perform a more fair comparison.
We follow the approach of \citet{Lagos2025}, explained below.
Observationally, we have errors on our measured SFRs and stellar masses. These can scatter to both higher or lower values. However, at high-z these will not scatter equally as we have significantly fewer quiescent galaxies than SF galaxies, and similarly fewer massive galaxies than lower-mass ones. This means that by including an extra observational error on SFRs and stellar masses from the simulations, we are likely to increase the observed number densities. In the practice, we add gaussian-distributed errors centred on 0 (meaning we assume no bias in our measurements) with a \sig of 0.3 dex to all stellar masses and SFRs. 
The sSFR selection criteria is then applied to the resulting quantities. This yields more galaxies making the quiescent cut.
The number densities computed from this approach are shown in the lower panel of Fig. \ref{fig:number-density-10}.

We can see that adding random errors to stellar masses and SFRs changes the number densities from simulations significantly, overall leading to an increase. This increase for the most part is still insufficient for most simulations, which remain below our measurements. In the case of \shark, adding errors leads to better agreement with the observations.

\begin{table}
    \caption{Number densities}
    \label{tab:num_dens}
    \centering
    \begin{tabular}{c|cccc}
        \hline
        \noalign{\smallskip}
        $z$ & $\rm M_*>10^{9.5}M_\odot$ & $\rm M_*>10^{10}M_\odot$ \\
        & ($\rm \times10^{-5}\ Mpc^{-3}$) & ($\rm \times10^{-5}\ Mpc^{-3}$) & $\rm N_{>9.5}$ & $\rm N_{>10}$ \\
        \noalign{\smallskip}
        \hline
        \noalign{\smallskip}
        2.0--2.5 & ${26.02}_{-2.40}^{+2.41}$ & ${22.63}_{-2.14}^{+2.14}$ & 377 & 328 \\
        2.5--3.0 & ${10.67}_{-1.36}^{+1.36}$ & ${8.93}_{-1.19}^{+1.19}$ & 154 & 129 \\
        3.0--3.5 & ${8.49}_{-1.26}^{+1.23}$ & ${6.64}_{-1.04}^{+1.04}$ & 119 & 93 \\
        3.5--4.0 & ${3.58}_{-0.78}^{+0.74}$ & ${2.61}_{-0.59}^{+0.62}$ & 48 & 35 \\
        4.0--4.5 & ${1.96}_{-0.58}^{+0.54}$ & ${1.49}_{-0.47}^{+0.44}$ & 25 & 19 \\
        4.5--5.0 & ${0.99}_{-0.37}^{+0.37}$ & ${0.50}_{-0.28}^{+0.24}$ & 12 & 6 \\
        5.0--6.0 & ${0.27}_{-0.13}^{+0.15}$ & ${0.14}_{-0.08}^{+0.09}$ & 6 & 3 \\
        6.0--7.5 & ${0.07}_{-0.06}^{+0.06}$ & ${0.03}_{-0.04}^{+0.05}$ & 2 & 1 \\
        \noalign{\smallskip}
        \hline
    \end{tabular}
    \tablefoot{Observed number densities for different mass ranges in redshift bins from $z=2$ to $7.5$.}
\end{table}

\section{Stellar mass functions}

\label{s.smfs}

\begin{figure}
    \centering
    \includegraphics[width=\linewidth]{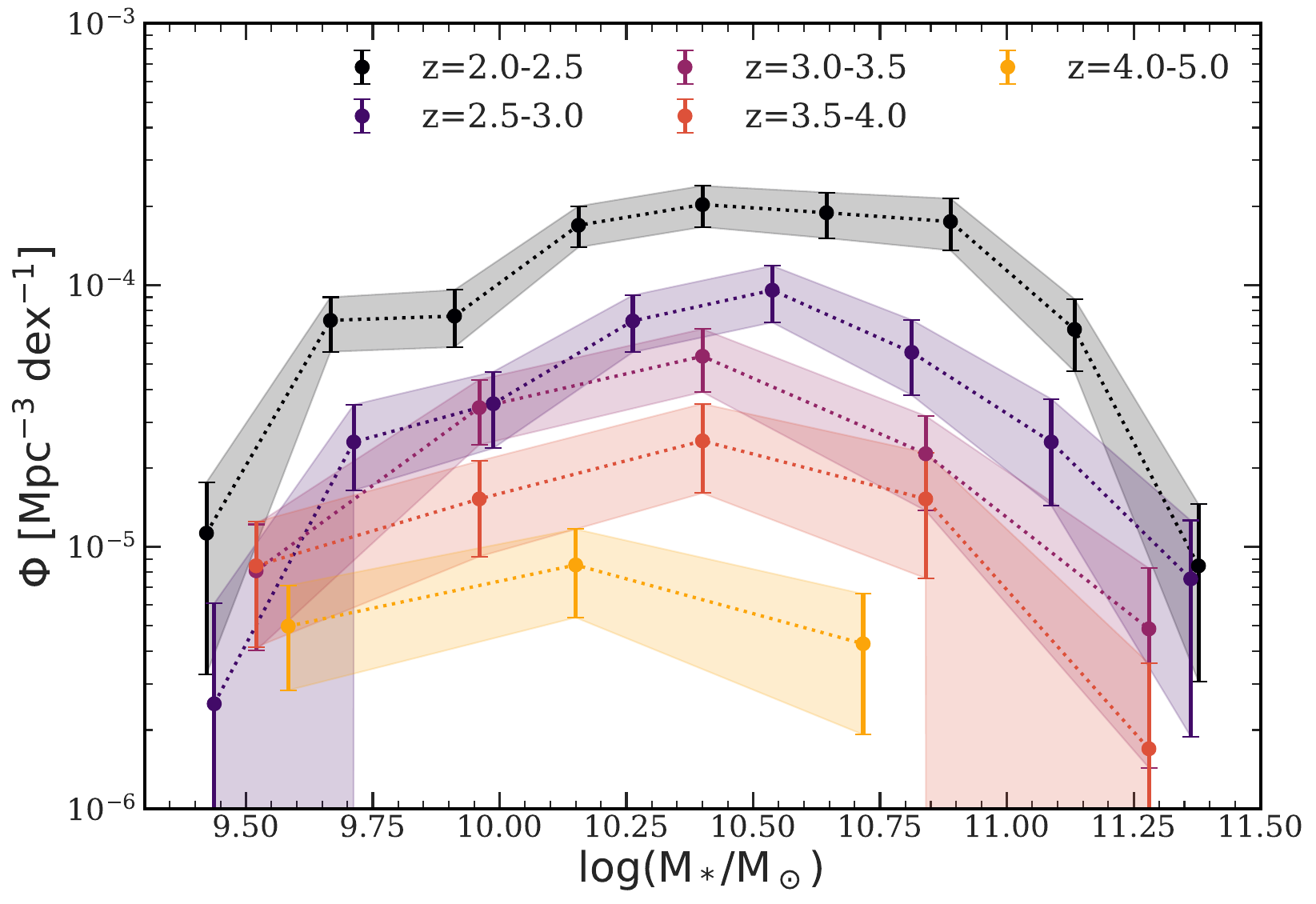}
    \includegraphics[width=\linewidth]{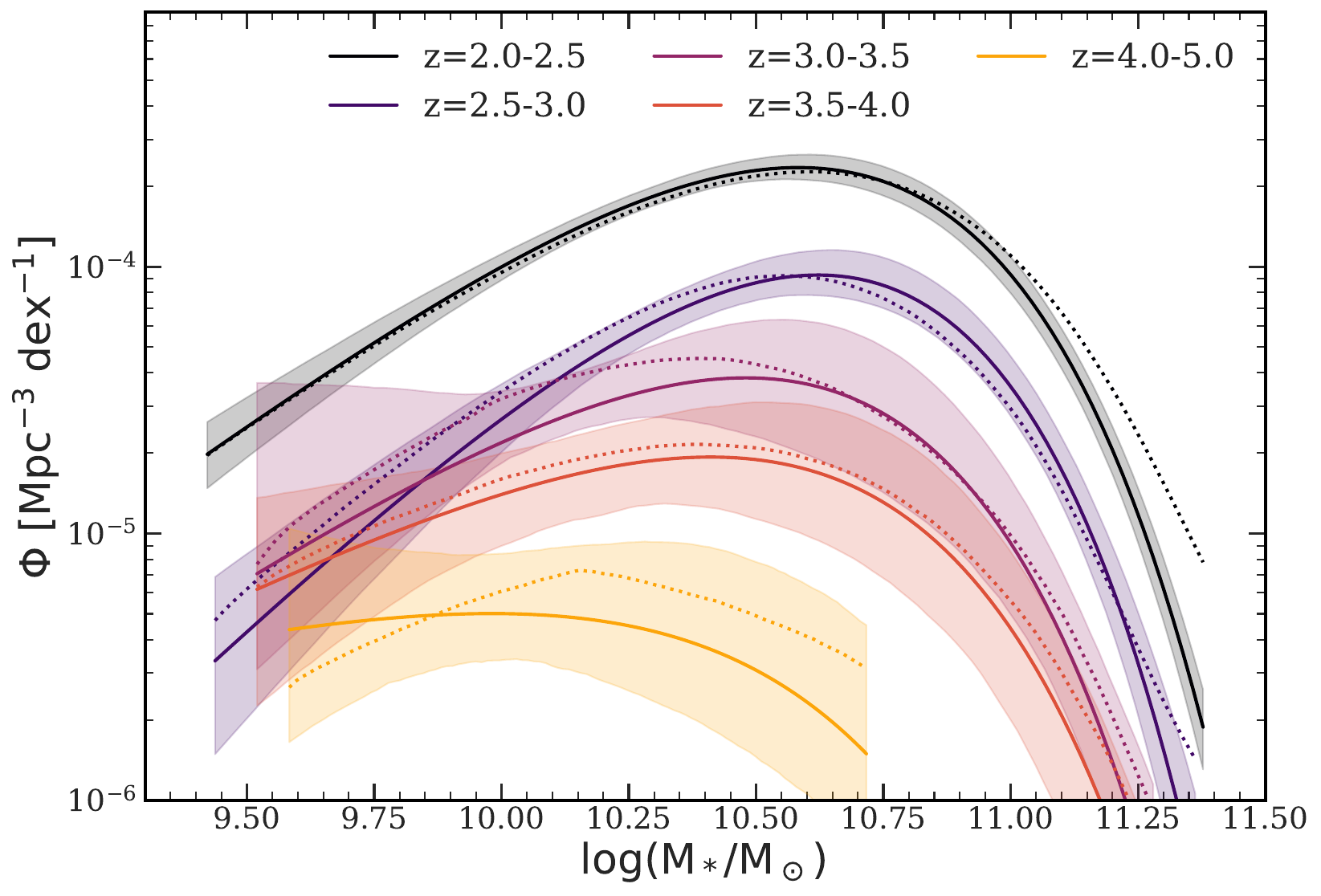}
    \caption{Upper: observed stellar mass function ($\Phi$) versus stellar mass for massive quiescent galaxies in bins of redshift. Lower: Fitted stellar mass function ($\Phi$) versus stellar mass for massive quiescent galaxies in bins of redshift. The solid line is our best-fit stellar mass functions including the correction for Eddington bias, the dotted lines exclude the Eddington bias correction. {The dotted lines show that we would significantly overestimate the number of most massive quiescent galaxies at $z=2-2.5$ without the correction.}}
    \label{fig:smf_observed}
\end{figure}

In addition to simply counting the number density of massive quiescent galaxies, we can also explore their mass distribution. To do this, we compute the stellar mass function (SMF) in various bins of redshift. A limitation of previous studies using JWST at these redshifts is their respective survey volume; they have not been large enough to compute a SMF for just the  quiescent galaxies. Even with our volume size, we cannot go above $z=5$, but we can explore the SMF from $z=2-5$. This enables a direct comparison to cosmological simulations and SAMs, albeit with the usual observational caveats (see Sec. \ref{app.s.smf_fitting}). 

In our case a key advantage over similar SMF works \citep[e.g.][]{McLeod2021,Weaver2023}, is that our  {entire sample is based upon a more stringent selection criteria for obtaining massive quiescent galaxies. We also have the benefit of JWST based data which enables us to fully probe the rest optical to high sensitivity.}
This means that we have more flexibility in our selection and, due to fewer targets, we can use Bayesian based SED modelling and visual inspection to remove contaminants. 

Fig. \ref{fig:smf_observed} shows the observed stellar mass function for our sample of high-z massive quiescent galaxies. We compute the stellar mass function for ranges of $z=2.0-2.5$, $z=2.5-3.0$, $z=3.0-3.5$, $z=3.5-4.0$ and finally $z=4.0-5.0$. 
We bin the masses in bins from $10^{9.3}M_\odot$<\Mstar<$10^{11.5}M_\odot$. 
We reduce the number of mass bins for the highest-z ranges due to a lack of very massive galaxies at those redshifts. 

We see from Fig. \ref{fig:smf_observed} that as z increases we obtain fewer massive quiescent galaxies of all masses (as expected from our number densities results), but also that this is more pronounced for masses closest to the turn over in the SMF (i.e. in the range $10^{10.3}M_\odot<M_*<10^{11}M_\odot$). The least amount of evolution is seen for the highest and lowest mass bins.
We find that, as has been seen before \citep[e.g.][]{McLeod2021, Weaver2023}, there are fewer low-mass and most-massive quiescent galaxies. 

In addition, we find a sharp jump in the growth at all masses from $z=4.0-5.0$ to $z=3.5-4.0$, mirroring what has been found previously for the overall galaxy population \citep{Weibel2024} and for the dusty star-forming galaxy sub-population \citep{Gottumukkala2024}. This suggests that during this epoch galaxies could be growing particularly efficiently.
We see that the growth in number densities with redshift does appear to have a mass-dependence, with the lower-mass quiescent galaxies experiencing less of an increase with redshift compared to other masses.
The next stage is to fit a functional form to the stellar mass function. This will enable a better understanding of its evolution and physics.

\subsection{Fitting the SMF}
 The functional form we will use to fit the stellar mass function is the single Schechter function \citep{Schechter1976}. This function has been shown to well reproduce stellar mass functions of single galaxy populations \citep[e.g.][]{Davidzon2017, Adams2021}.
The Schechter function is defined as follows
\begin{equation}
\rm
    \Phi(M)\ dM = \Phi^*\left(\frac{M}{M^*}\right)^\alpha \exp\left(-\frac{M}{M^*}\right)d\left(\frac{M}{M^*}\right)
\end{equation}
where $\rm M^*$ is the characteristic stellar mass that corresponds to the "knee" of the SMF, the point at which the exponential cutoff and powerlaw slope start. $\Phi^*$ is the normalisation and corresponds to the number density at $\rm M^*$, and $\alpha$ is the strength of the powerlaw slope.
Fitting the SMF with a common functional form enables easy comparison between different works and their parameters. { It also provides a simple model that can be linked to theory.}

However, before we fit the functional form, we should explore the relative uncertainties involved in fitting a SMF. 
This is explored in depth in Appendix \ref{app.s.smf_fitting}. For now we note that as we are probing massive quiescent galaxies we can detect down to our mass cut. We take into account cosmic variance following the methodology of \citet{Jespersen2025}. Finally, we correct for \citet{Eddington1913} bias via convolving the SMF with a Voigt profile.

In order to fit the SMF we use the code \textsc{DYNESTY} \citep{Speagle2020} which uses Dynamic Nested Sampling \citep{Skilling2004, Higson2019} in order to explore the posterior space for the SMF. This approach better enables degeneracies between various quantities to be explored, whilst also being able to deal with a large number of dimensions within the parameter space. 
We fit for $\phi^*$, $\rm M^*$ and $\alpha$ with the log likelihood calculated for a simple Gaussian error model \citep[e.g. $\chi^2$, following][]{Hogg2010}.
We use uniform priors on all parameters. 

The resulting best-fit stellar mass functions are shown in Fig. \ref{fig:smf_observed}. The errors are obtained from the 16th and 84th percentiles of the resulting distribution of the stellar mass functions computed using the posteriors of the three parameters. This has the effect of propagating through the uncertainties on $\phi^*$, $\rm M^*$ and $\alpha$.  

The exact values for the best-fit SMFs are reported in Table \ref{tab:smf}.
These show that the "knee" of the SMF ($M^*$) remains around $10^{10.3-10.5}$ regardless of redshift, hence it does not show signs of evolution from $z=2-5$. 
The "knee" of the stellar mass function is classically used to explain the switch between high and low-mass quenching mechanisms \citep[alongside the switch from a single Schechter to double Schechter function required by environmental quenching][]{Peng2010, Shuntov2025}. The idea is that the powerlaw slope set by $\alpha$ describes primarily the lower mass quenching mechanism, whilst the exponential cutoff describes the higher-mass quenching mechanism (typically thought to be AGN feedback). Therefore, the knee describes the mass range at which this switchover occurs.
This suggests that the switch between high and low-mass quenching mechanisms in the quiescent galaxy SMF does not appear to vary significantly with redshift.

We see an expected decrease in the normalization of the Schechter function which reiterates our findings with the number densities. We see fewer massive quiescent galaxies of every mass at higher redshift rather than simply seeing a different distribution.
However, we do see some tentative evidence for a flattening of the power-law slope $\alpha$ (although this is particularly difficult to constrain). This could suggest that the lower mass quenching mechanism is evolving less with redshift compared to the higher mass mechanism. Alternatively, the greater high-mass build up with redshift may correspond to growth via major or minor mergers \citep{Puskas2025}, however, recent works have suggested a lack of major mergers among high-z massive quiescent galaxies
\citep{Baker2025a,D'Eugenio2024, Pascalau2025, Chittenden2025}.

\begin{table}
      \caption[]{Best-fit stellar mass function parameters}
         \label{tab:smf}
         \begin{tabular}{c|ccc}
            \hline
            \noalign{\smallskip}
                   $z$ & $\rm M^*$ & $\phi^*$ & $\alpha$ \\

                    & $\rm log(M_\odot$) & $\rm \times 10^{-5}\ Mpc^{-3}$ & \\
            \noalign{\smallskip}
            \hline
            \noalign{\smallskip}

           $z=2.0-2.5$ & ${10.43}_{-0.03}^{+0.03}$ 
                & ${25.68}_{-2.26}^{+2.11}$
                & ${1.42}_{-0.15}^{+0.17}$ \\
                $z=2.5-3.0$ & ${10.35}_{-0.07}^{+0.07}$
                & ${8.20}_{-2.16}^{+1.51}$
                & ${1.86}_{-0.38}^{+0.45}$ \\
                $z=3.0-3.5$ & ${10.37}_{-0.13}^{+0.28}$
                & ${4.36}_{-2.39}^{+1.56}$
                & ${1.28}_{-1.22}^{+0.72}$ \\
                $z=3.5-4.0$ & ${10.42}_{-0.15}^{+0.15}$
                & ${2.28}_{-1.13}^{+0.99}$
                & ${0.97}_{-0.61}^{+0.78}$ \\
                $z=4.0-5.0$ & ${10.34}_{-0.28}^{+0.96}$
                & ${0.48}_{-0.42}^{+0.55}$
                & ${0.44}_{-1.01}^{+1.29}$ \\
    
            \noalign{\smallskip}
            \hline
         \end{tabular}
         \tablefoot{The best-fit \citet{Schechter1976} function parameters for each SMF fit for each redshift}
   \end{table}

\subsection{Comparison to simulations}

\begin{figure*}
    \centering
    \includegraphics[width=0.99\columnwidth]{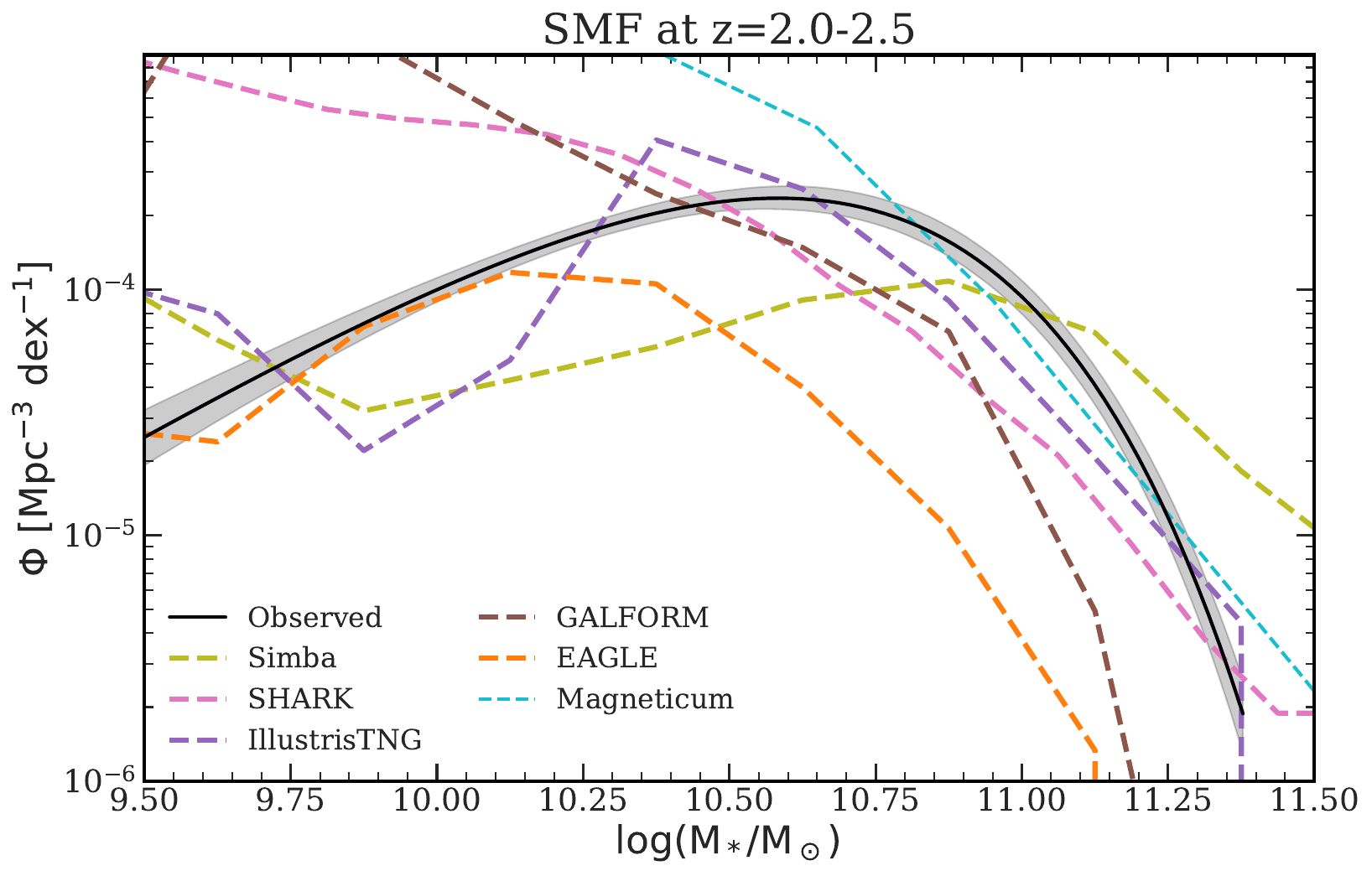}
    \includegraphics[width=0.99\columnwidth]{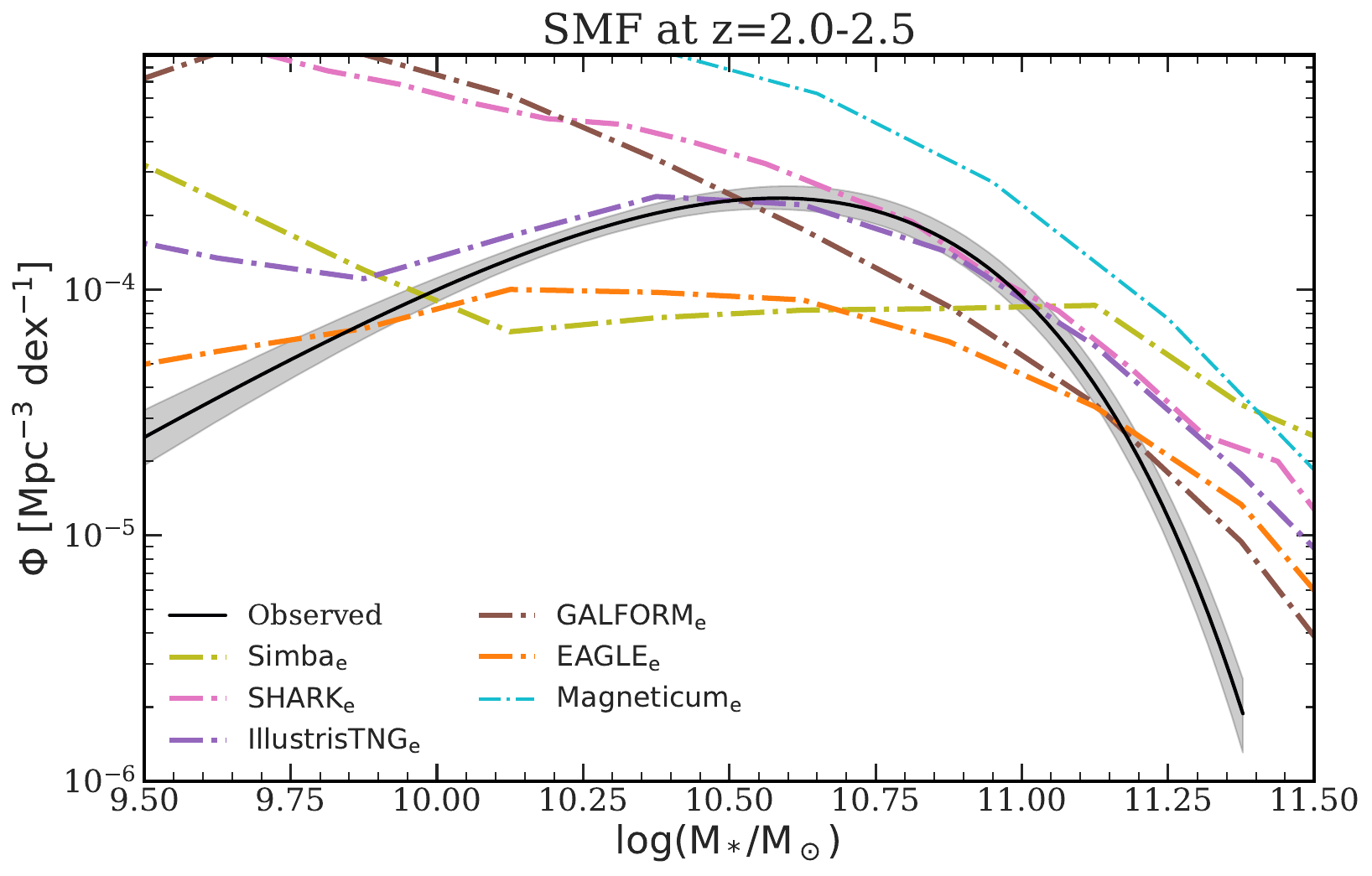}
    \includegraphics[width=0.99\columnwidth]{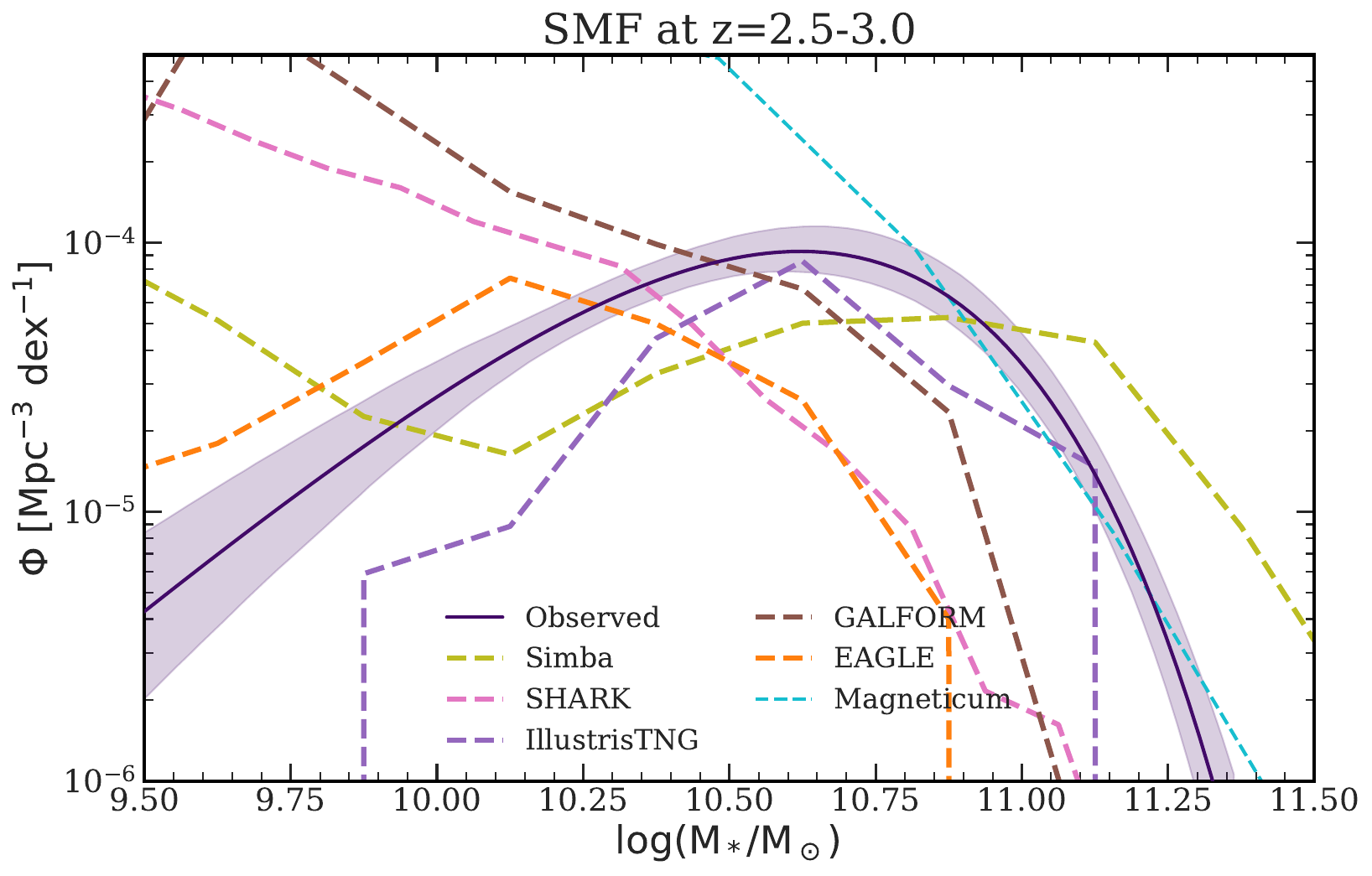}
    \includegraphics[width=0.99\columnwidth]{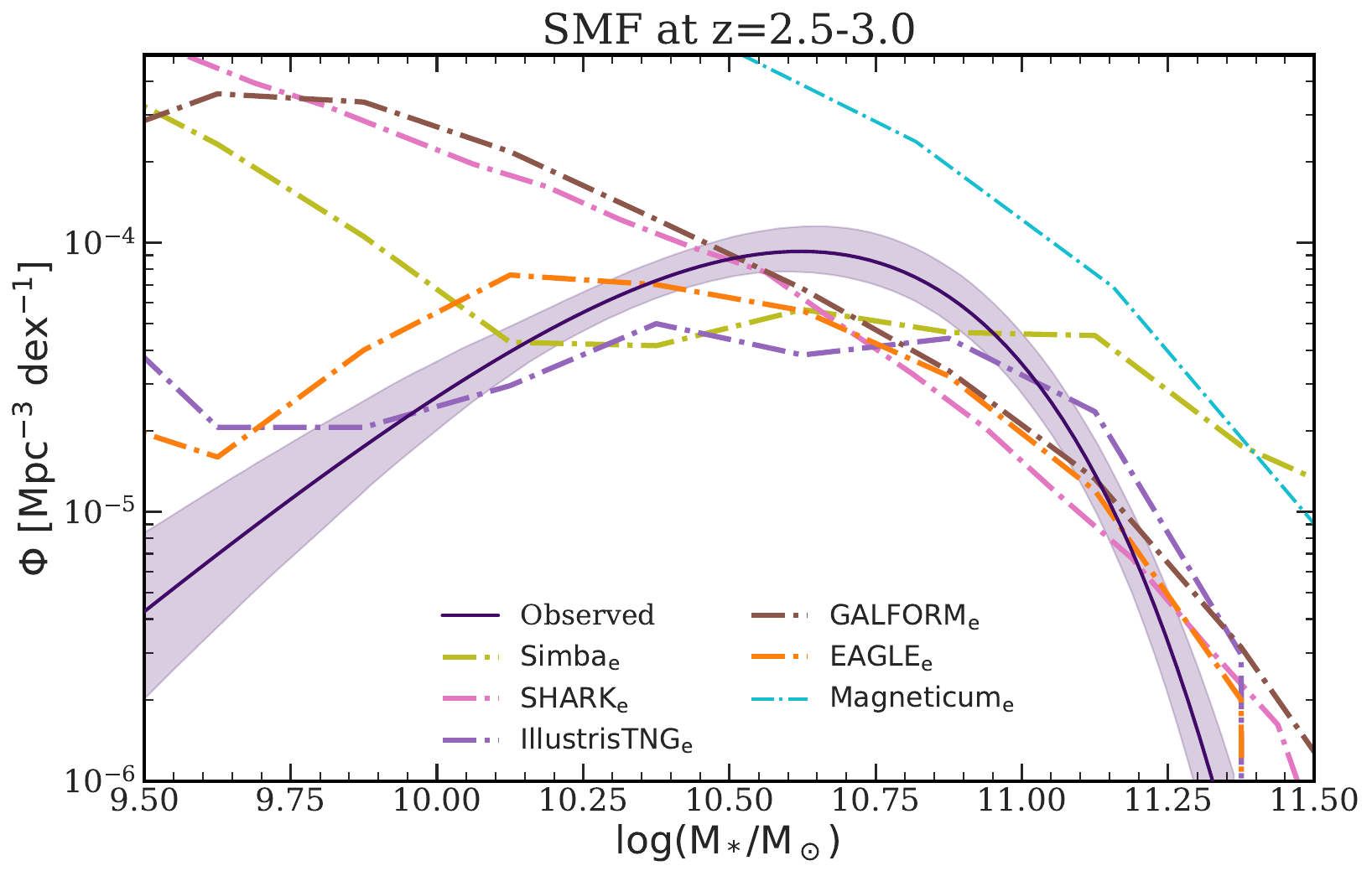}

    \includegraphics[width=0.99\columnwidth]{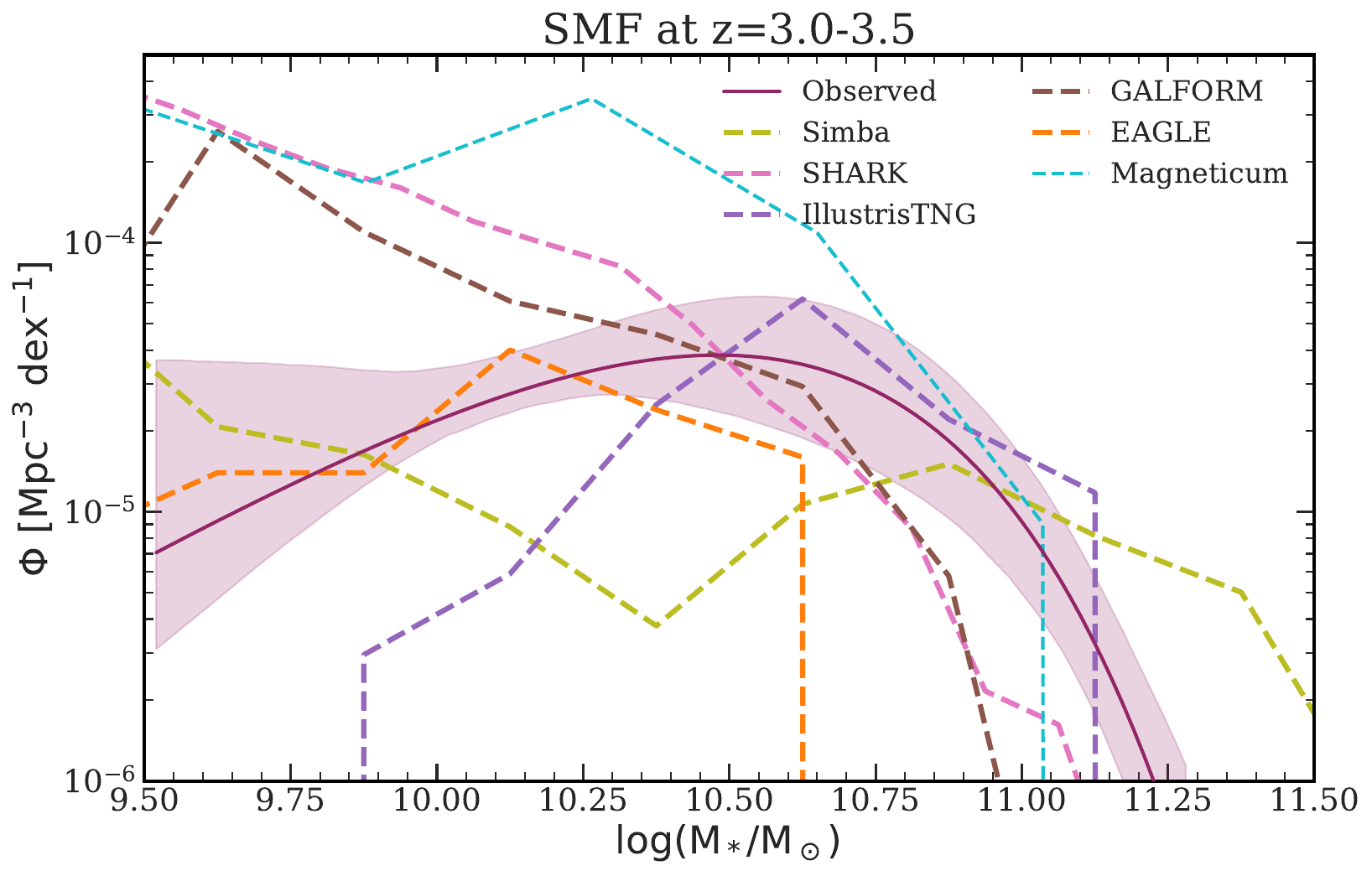}
    \includegraphics[width=0.99\columnwidth]{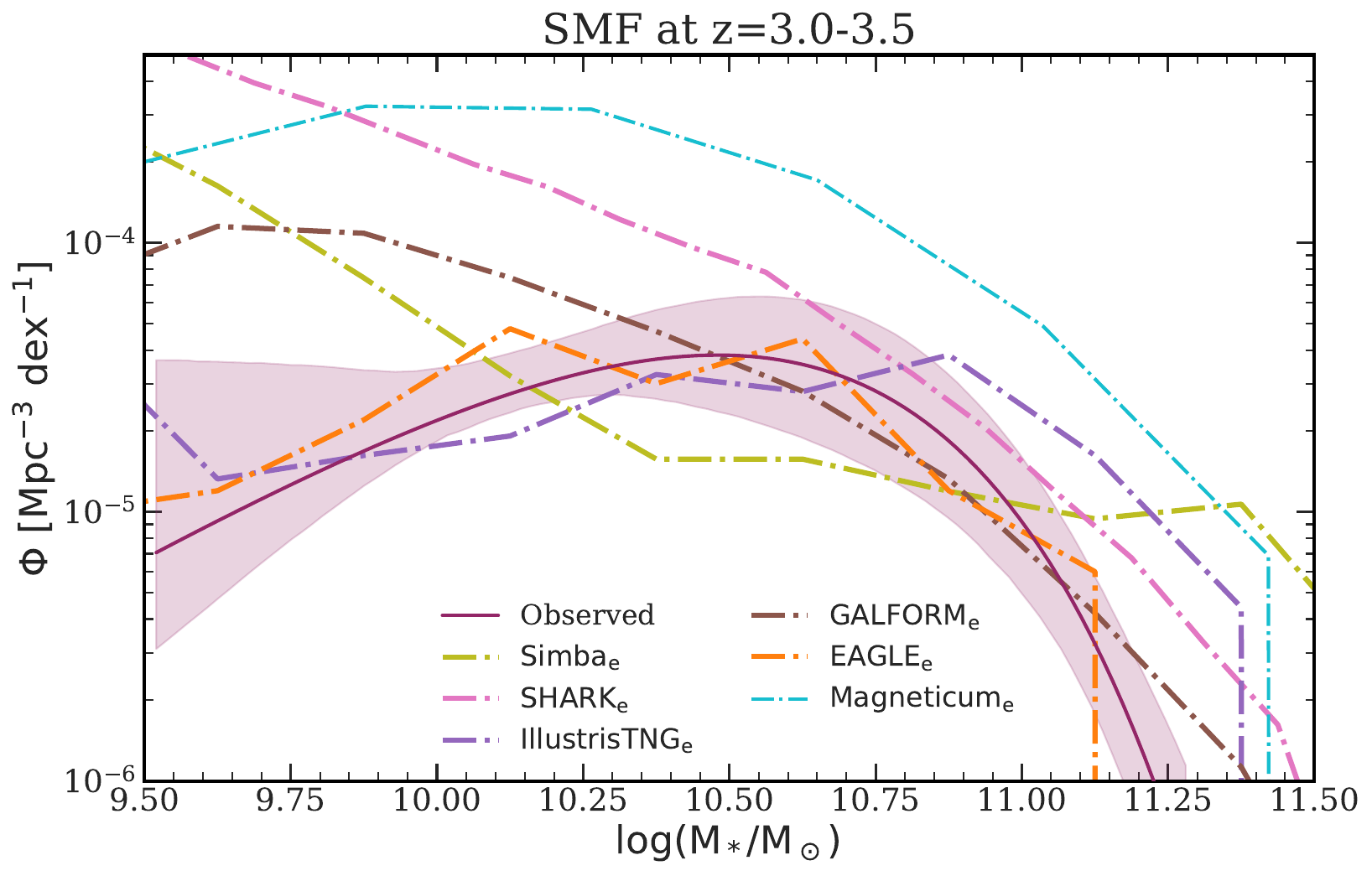}
    \includegraphics[width=0.99\columnwidth]{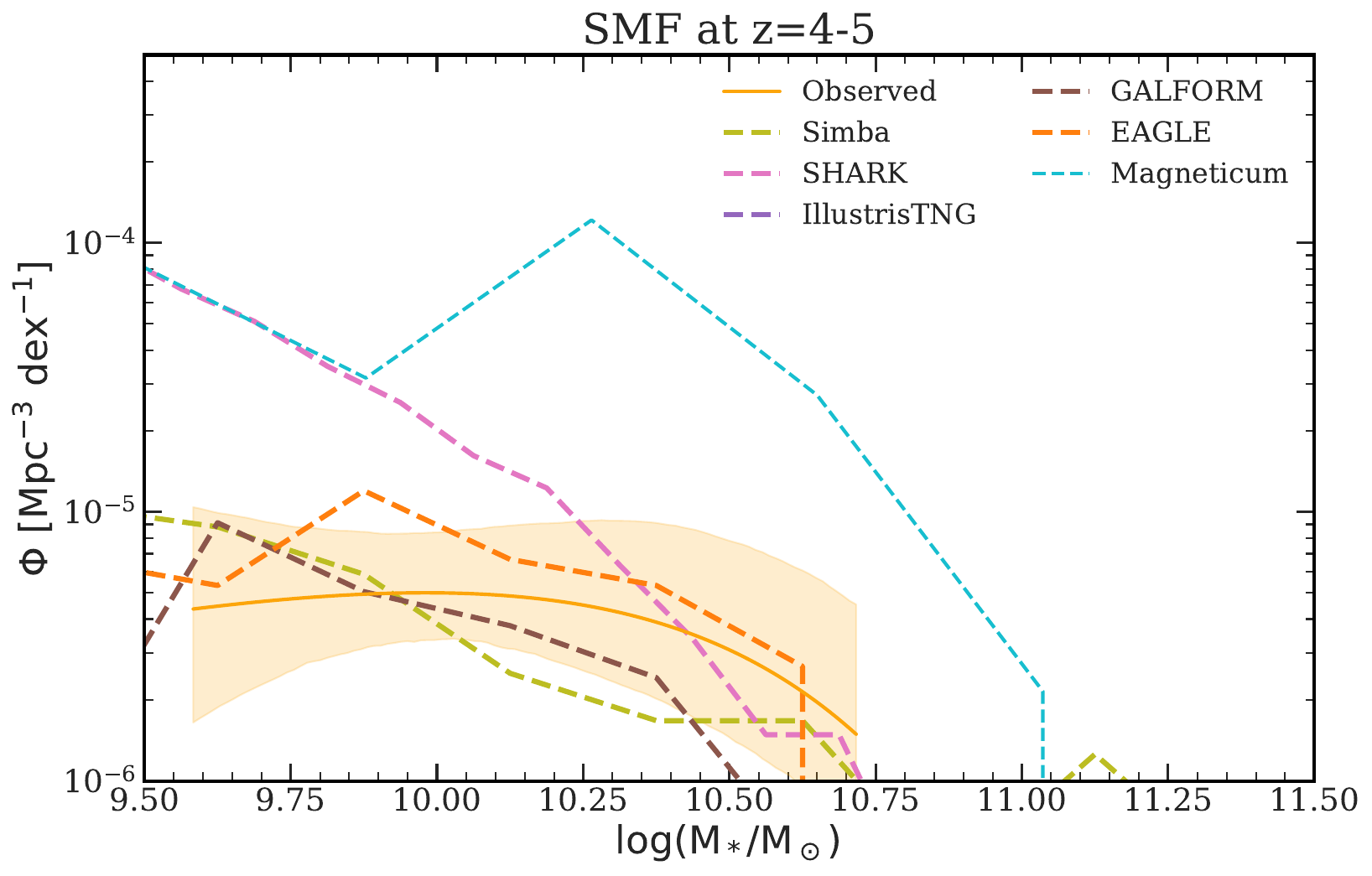}
    \includegraphics[width=0.99\columnwidth]{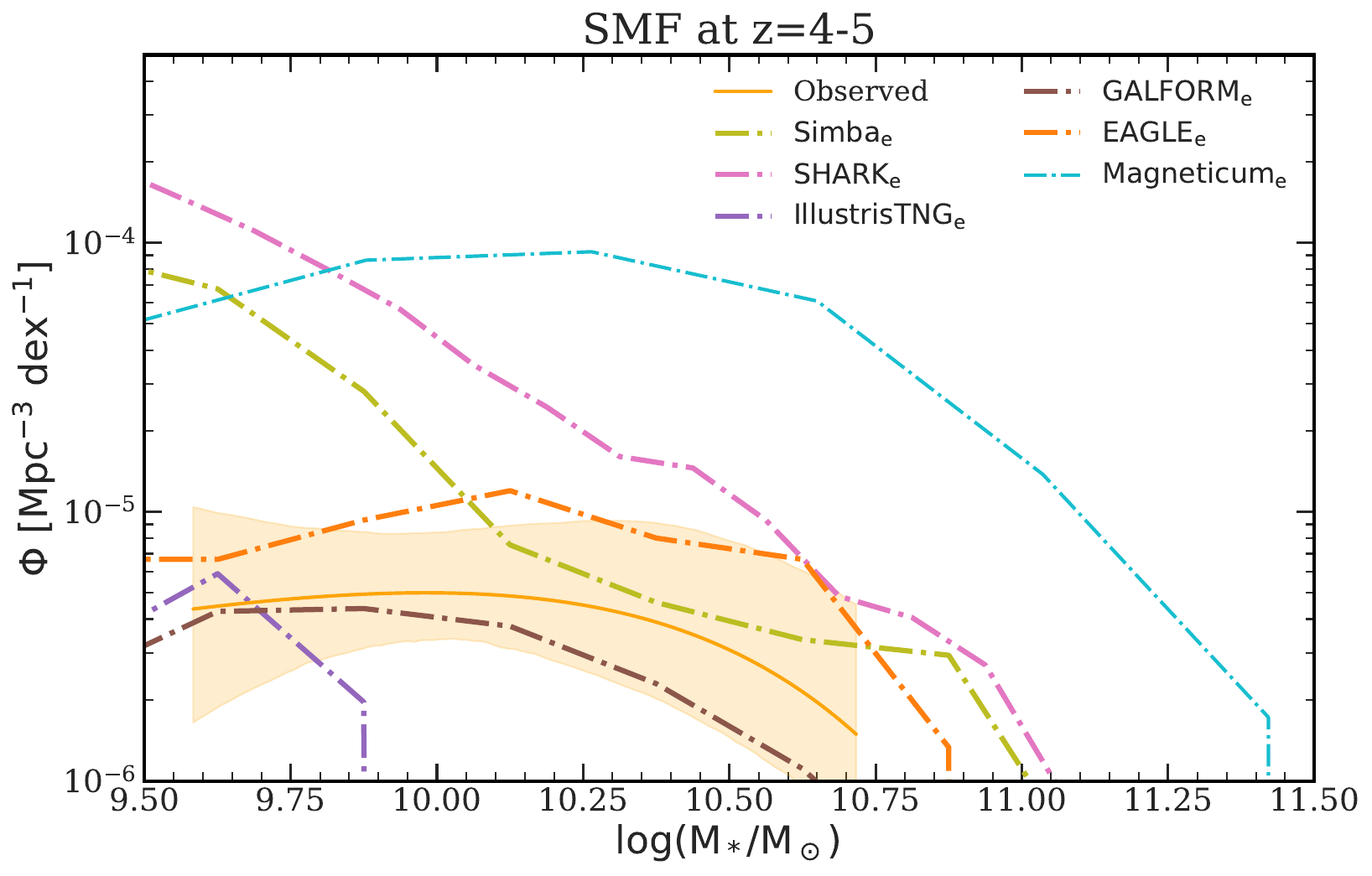}
    
    \caption{Comparison between our observed stellar mass function (solid lines) at $z=2-2.5$ (upper panel) $z=2.5-3$ (upper middle panel), $z=3-3.5$ (lower middle panel) and $z=4-5$ (lower panel), alongside 6 cosmological simulations denoted by dashed lines  \citep[left hand panels,][]{Lagos2025}. The right hand panel shows the same, but with Gaussian scatter added to the stellar masses and SFRs for the 6 simulations. We see that the simulations cannot accurately reproduce the observed SMFs. }
    \label{fig:smf_sim}
\end{figure*}

The real power of exploring stellar mass functions for quiescent galaxies comes from comparing them to cosmological simulations and SAMs. 
This is more helpful in our case as we have a larger more robust observational sample, that has been selected primarily based on the same sSFR criteria as from the simulations enabling a much more like-for-like comparison.

We compare with the stellar mass functions from \eagle, \tng, \simba, \shark, \galform, and \magneticum in various bins from $z=2-5$.
In order to match the observed bins of our observation we average over the available simulation snapshots' SMF within that bin. 
The results are shown in the left panel of Fig. \ref{fig:smf_sim}. Our observed SMF is given by the solid lines and shaded region, whilst the results from the simulations is given by the dashed line. 
We can see immediately that none of the cosmological simulations or SAMs fully reproduce the observed SMF for the massive quiescent galaxies at any redshift probed in this work. As was found by \citet{Lagos2025} for just $z=3$, some simulations and SAMs over produce lower mass quiescent galaxies and under produce higher mass, whilst others fail at a variety of different mass scales. 

There is a large disagreement between the simulations themselves on the shapes of the quiescent galaxy SMFs.  {These shapes are a convolution of the total SMFs in the simulations (which tend to be more power-law-like) with the passive fractions as functions of stellar mass \citep[which rarely look to be smooth functions, e.g.][]{Weaver2023}. This makes them complicated to understand physically.} However, a significant cause of the disagreement is likely due to the various feedback implementations within the simulations \citep[][]{Lagos2025}.
This shows that even SAMs such as \shark, which accurately reproduce the observed number densities for \Mgten, cannot reproduce the observed mass distribution.

In the right-hand panel of Fig. \ref{fig:smf_sim} we compare to the simulations with the addition of the gaussian errors on the observed quantities (stellar mass and SFR). These are denoted with the dot-dash lines. Adding errors has the effect of broadening the SMF, particularly helping to reproduce the higher-mass end, in the cases where the models struggle with producing enough high-mass galaxies, but there are still significant differences with the observations. 
This is again most apparent at the lower mass end, with \eagle being the closest to reproducing the SMF there, but largely failing at the high mass end. {It is also worth recalling that all these simulations failed to reproduce the observed number densities in Sec. \ref{s.number_densities10}.}

\subsection{Field-to-field and colour selection variations}

\begin{figure*}
    \centering
    \includegraphics[width=1.\linewidth]{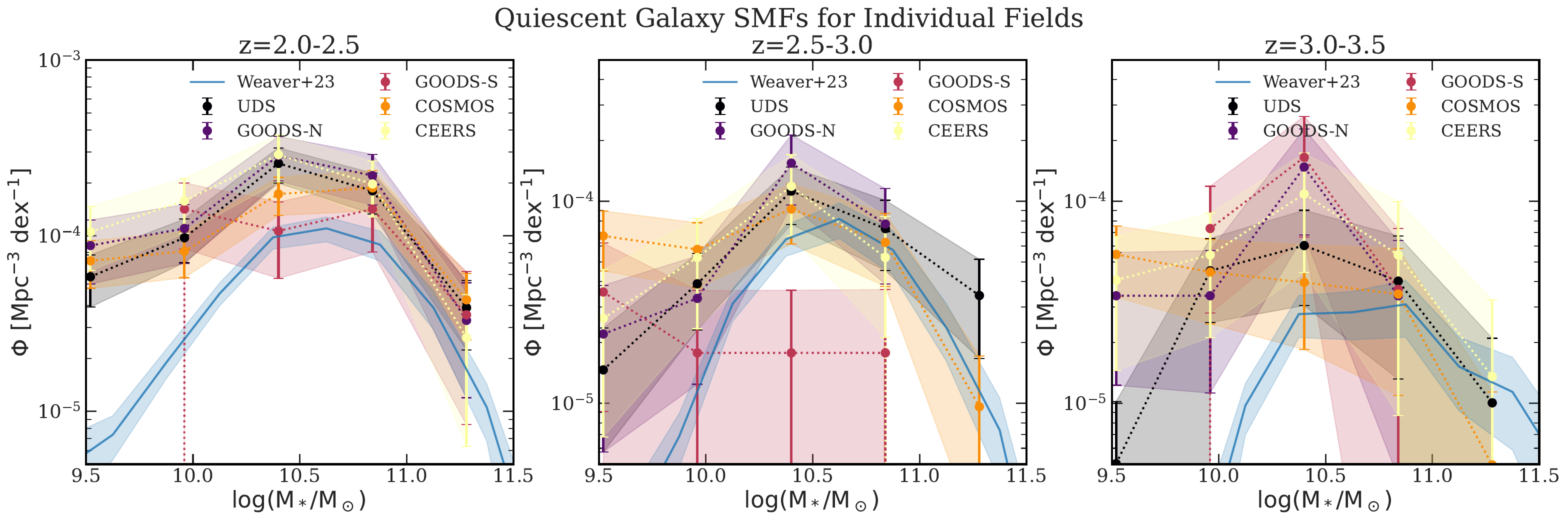}

    \includegraphics[width=1\linewidth]{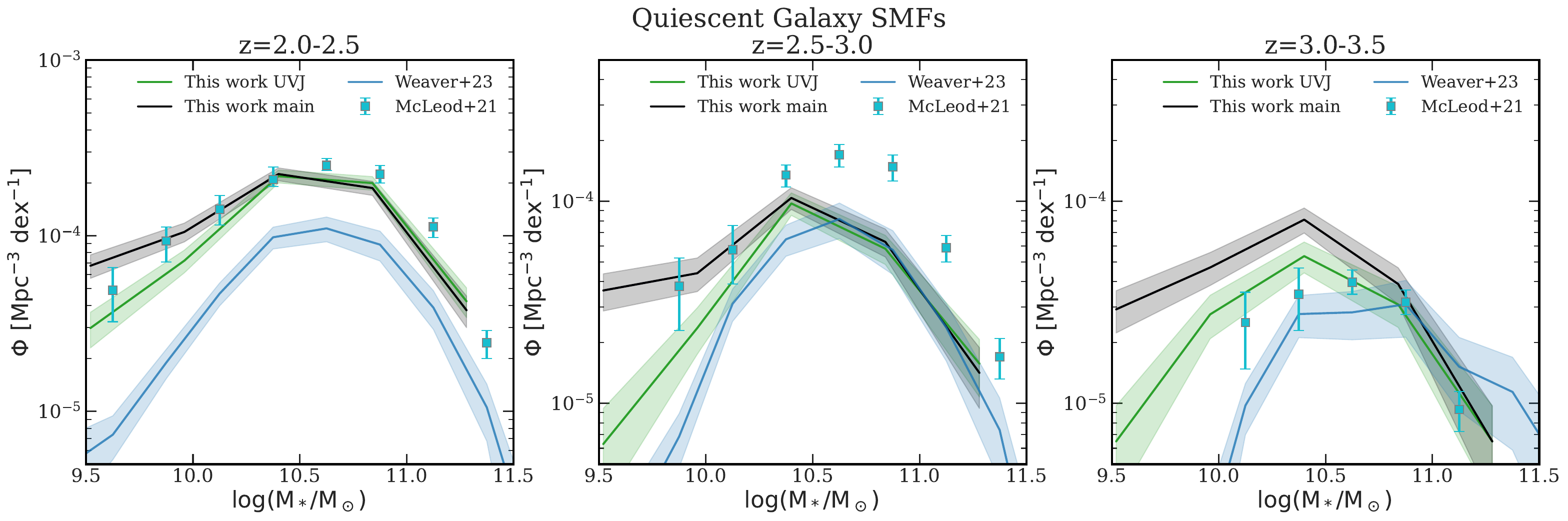}

    \caption{Upper panels: field-to-field variations in the SMF with the same selection criteria as previously. We also plot the fitted SMF from \citet{Weaver2023} in blue for their quiescent galaxy sample. Left for $z=2-2.5$, middle for $z=2.5-3.0$ and right for $z=3.0-3.5$. We see significant field-to-field variations underlining the effects of cosmic variance in small field observations. This increases at higher redshifts.
    Lower panels: the same but using a stricter UVJ cut to better replicate the selection criteria of previous works \citet{McLeod2021,Weaver2023}. {The observed quiescent galaxy SMFs from this strict UVJ selection are denoted by the green line.} We find greater consistency with the \citet{Weaver2023} and \citet{McLeod2021} results, highlighting the importance of selection criteria in obtaining as many quiescent galaxies as possible. {We plot the SMF for the sample with our main selection criteria as the black line, showing how the largest difference between the selection criteria is for low-mass quiescent galaxies and at higher redshifts}.}
    \label{fig:smf_field}
\end{figure*}

Due to our use of several different fields in different regions of the sky, we can investigate possible field-to-field variations in the SMF. 
We only compute this from redshift $z=2.0-2.5$, $z=2.5-3.0$, and $z=3-3.5$ in order to ensure we have enough galaxies for our bins. The results of this are shown in Fig. \ref{fig:smf_field}.
The upper panel shows the SMF computed on individual fields, in this case PRIMER-UDS, PRIMER-COSMOS, CEERS-EGS, and GOODS. 
Also overplotted is the equivalent SMF from \citet{Weaver2023}. \footnote{ {In this work they used ground based photometry based on the full COSMOS field which corresponds to an area of 1.27 square degrees. The advantage of that work is the significantly larger volume than our $\rm \sim 800 arcmin^2$. However, the ground based imaging naturally has reduced depth (approaching only 26 mag in H band) and it lacks the longer wavelength filters we have access to (which enable a better probe of the rest-frame optical). In addition, by using a combination of different fields in different regions on the sky we benefit from multiple different sightlines, helping reduce the effect of cosmic variance compared to a single large field.}}

In Fig. \ref{fig:smf_field}, we can immediately see that there is a natural variation in the stellar mass functions for each field, particularly at the low-mass end.
CEERS clearly has the greatest number of quiescent galaxies for its area at the low-mass end, which mirrors previous results finding an overabundance of massive quiescent galaxies within this field \citep{Jin2024,Ito2025,Valentino2023}. 
Meanwhile, GOODS-S varies significantly depending on masses. This is showing that cosmic variance does have an effect, particularly on small survey areas, hence one should take into account multiple different fields with different positions on the sky when computing number densities and SMFs for massive quiescent galaxies.

For the $z=2.5-3.0$ bin we see much more variation, showing the increasing impact of cosmic variance at higher-z when sources become rarer. This becomes even more pronounced at $z=3.0-3.5$, where we can see clear differences between the shapes and values of the SMFs between the various fields.
It should also be noted that we are probing the redshift ranges that should be least affected by cosmic variance or Poisson counting statistics (i.e. $z<4$) and that the effect is only likely to be more significant at higher redshifts ($z>4$).
Therefore, this provides strong evidence that cosmic variance will play a role for individual fields, as we clearly see this effect at $z = 2.5-3.5$.
We also see that for the $z=2-2.5$ bin in particular, we see many more lower mass massive quiescent galaxies than were found in \citet{Weaver2023}. We now explore what is driving this offset.

In the lower panel of Fig. \ref{fig:smf_field} we show SMFs computed with a much stricter UVJ selection. We use the \citet{Schreiber2015} criteria. Whilst not the same selection criteria as \citet{Weaver2023}, it is more strict than our sSFR criteria and should enable any variation to be seen. It is also very close to the UVJ criteria used in \citet{McLeod2021}.
There is a clear drop in the number of low-mass massive quiescent galaxies seen at $z=2-2.5$, $z=2.5-3$ and a clear drop in almost all masses at $z=3-3.5$. This can be seen by the offset between the SMFs for our full sample (black line) and the purely UVJ selected sample (green line). 
 {At $z=2-2.5$ we become more consistent with the results of \citet{McLeod2021}, whilst at higher redshifts we become more consistent with the results of \citet{Weaver2023}. This once again highlights the importance of a broad selection criteria (i.e. Bayesian based SED modelling, in addition to colours) in order to accurate uncover quiescent galaxies, particularly at lower masses and/or higher redshifts. 
This suggests that previous studies \citep[e.g.][]{McLeod2021,Weaver2023}, are missing significant numbers of lower-mass or higher-z quiescent galaxies through their selections.}

\subsection{Stellar mass density of massive quiescent galaxies}

\begin{figure}
    \centering
    \includegraphics[width=1\linewidth]{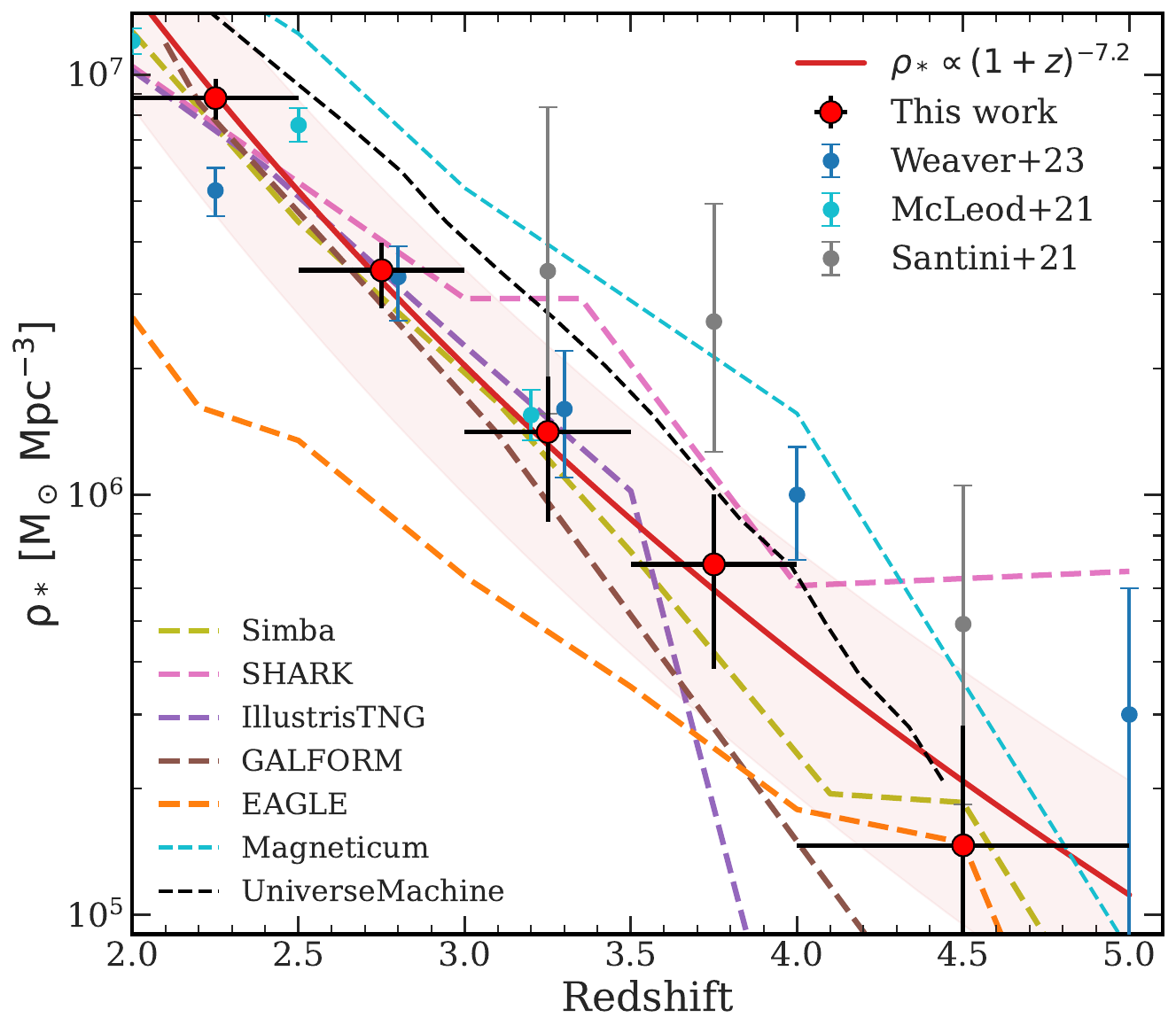}
    \caption{Stellar mass density (SMD) of quiescent galaxies versus redshift based upon our best-fit SMFs (red points) {with our best-fit overplotted (red line and shaded region)}. Also overplotted is a comparison for quiescent galaxies from \citet{Weaver2023}, blue points, \citet{Santini2021}, grey points, \citet{McLeod2021}, light blue points, and theoretical predictions from \textsc{UniverseMachine} \citep{Behroozi2019}, black dashed line. {We slightly offset the $z=3.25$ points for the \citet{McLeod2021} and \citet{Weaver2023} stellar mass densities in order to aid visibility within this figure.} In addition, we plot SMDs from \simba, \shark, \tng, \galform, \eagle and \magneticum obtained by integrating their SMFs. {We see a steeper increase in the observed SMD from $z=5$ to $z=2$ corresponding to $\rho_*\propto (1+z)^{-7.2\pm0.3}$, highlighting the increasing importance of galaxy quenching over this epoch.}}
    \label{fig:SMD}
\end{figure}

Another area of interest is exploring the stellar mass density (SMD) of a particular population of galaxies in the universe \citep[e.g. ][]{Muzzin2013,Madau2014}. This can be computed a number of ways (e.g. from the cosmic star-formation rate density), but in our case this is a straightforward matter of integrating the obtained stellar mass functions at each redshift. We use our fitted form of the stellar mass function with a commonly chosen lower mass limit of $M_*>10^8 M_\odot$ \footnote{We test varying this by integrating from $10^9 M_\odot$, as the simulations only go down to $10^9 M_\odot$, and find it makes no appreciable difference, as the very low-mass end is sub-dominant.}.
We integrate using our best-fit Schechter function for each redshift, extrapolating to lower masses than those explored in our SMFs.

Fig. \ref{fig:SMD} shows our stellar mass density versus redshift compared to other studies. It shows that we are seeing a steeper evolution in our SMD compared to other studies such as \citet{Weaver2023}. We find less mass per unit volume at higher redshifts and more mass at lower z. We note that we fully propagate through the errors from our fitted SMF. The increase in $\rho_*$ at $z=2.0-2.5$ is in keeping with our findings for the number densities and SMFs where we see an overabundance compared to other works. Interestingly, this is not the case at $z=2.5-3.0$ or $z=3.0-3.5$ which match the results of \citet{Weaver2023} (as can be seen in Fig. \ref{fig:smf_sim}).   
{We quantify the strength of the increase we see with a simple parametric model of the form $\rho_*=\rho_0(1+z)^\alpha$, where $\rho_0$ is a normalisation constant and $\alpha$ is the power-law index. We find that our best-fit model has $\rm \rho_0=4.1\times 10^{10}\ M_\odot\ Mpc^{-3}$ and $\alpha=-7.2\pm0.3$. This means that we are seeing an increase of the form $\rho_*\propto (1+z)^{-7.2\pm0.3}$ from $z=5$ to $z=2$. This is much higher than has been found in previous works, such as \citet{Muzzin2013}, who found $\rho_*\propto (1+z)^{-4.7\pm0.4}$.  }

We also compare our results to two other pre-JWST works, \citet{Santini2021} and \citet{McLeod2021}. We find reasonable agreement with \citet{McLeod2021}, particularly at $z=3.0-3.5$, but find lower mass per unit volume than \citet{Santini2021} (after correcting for IMF assumptions). 
{We again also compare with cosmological simulations and SAMs. We find that both \eagle and \magneticum dramatically fail to reproduce the observed SMD, with one severely underestimating and the other severely overestimating.} \tng matches it well up to $z=3.5$ where it dramatically falls (due to having no high-z massive quiescent galaxies). \shark overpredicts and \galform slightly underpredicts the SMD at $z>3.0$. 

Interestingly, \simba well reproduces the SMD. This is surprising as it failed to reproduce the observed number densities or SMF, but goes to show how difficult it is to accurately reproduce all of these within the same simulation. Clearly the shape of the SMFs in \simba is dramatically different to observations, but the area underneath remains the same.
We also overplot predictions from \textsc{UniverseMachine} \citep{Behroozi2019, Weaver2023} which slightly over predicts the SMD at each redshift, i.e. the normalisation is too high, but it matches the shape well.

As shown by the fit, the increase in the stellar mass density with redshift appears much larger in our work with $\rho_*\propto (1+z)^{-7.2\pm0.3}$. This suggests that galaxy quenching activity was significantly higher in those epochs than previously thought.

\section{High-z candidates}

\begin{figure}
    \centering
    \includegraphics[width=0.99\linewidth]{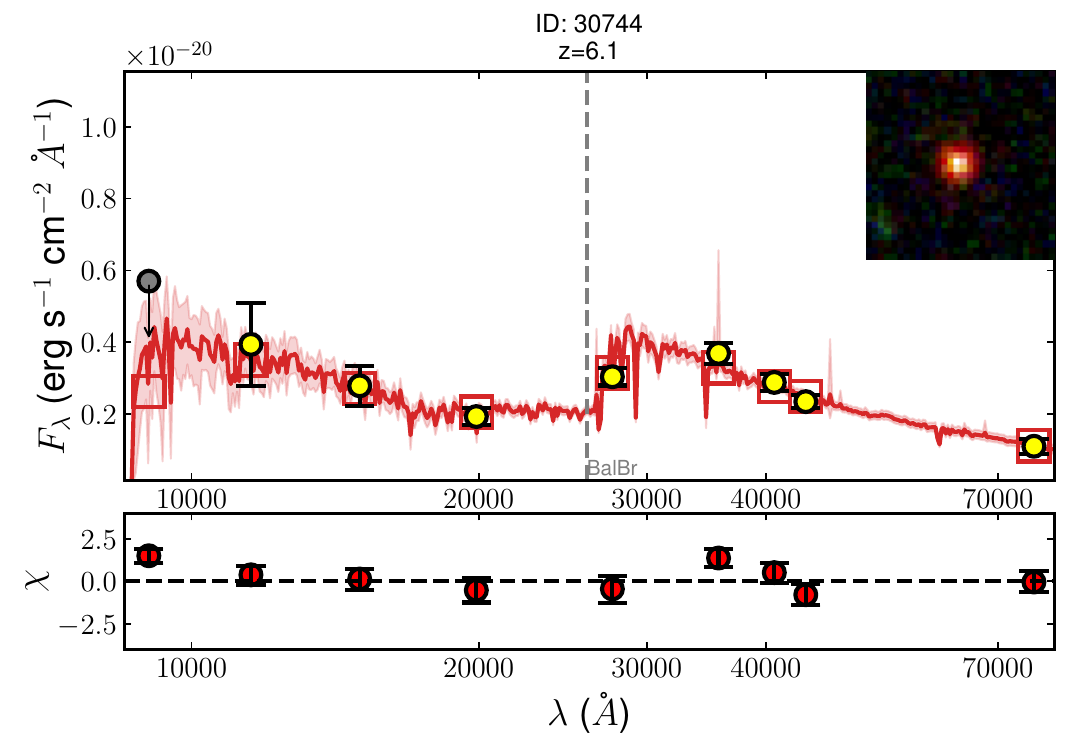}
    \includegraphics[width=0.99\linewidth]{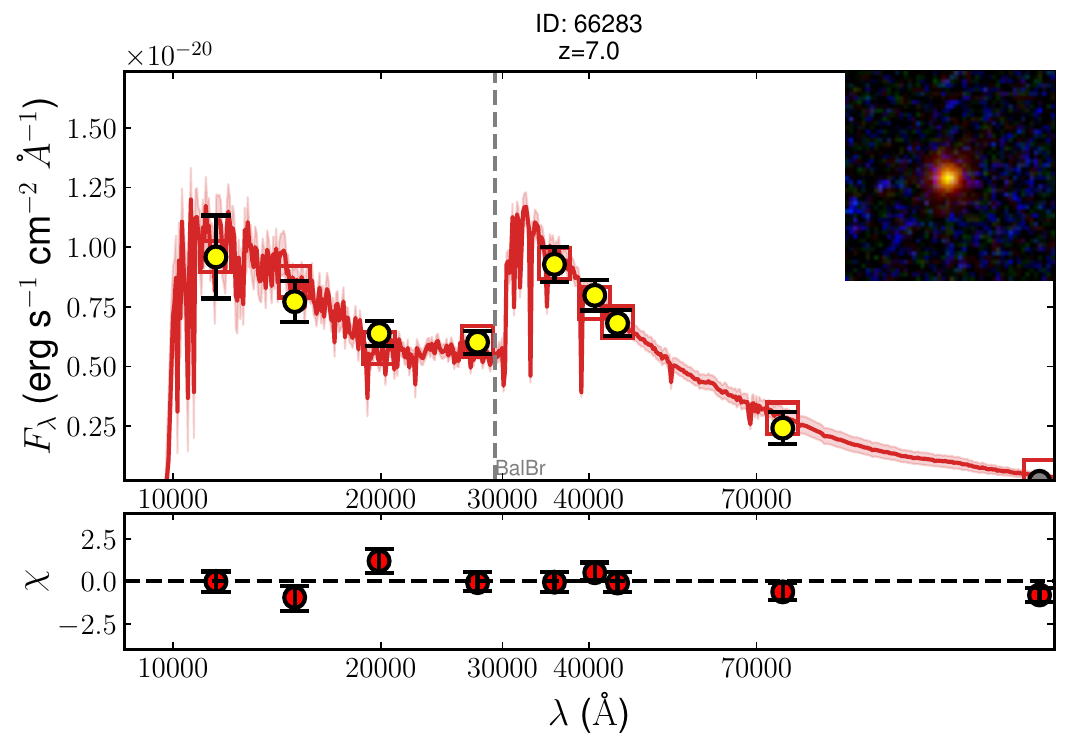}
    \caption{\pipes SED fits for our two highest redshift candidates \citep[the highest-z of which is confirmed by spectroscopy][]{Weibel2024qgal}. The yellow points are the observed photometry and the grey points correspond to non-detections with a SNR<3. The red-squares correspond to the best-fit model photometry and the red line corresponds to the best-fit model spectrum. Uncertainties in the best-fit model spectrum are denoted by the shaded regions which correspond to the 16th and 84th percentiles of the resulting distribution. An RGB image of the galaxy {corresponding to F115W-F200W-F444W} is included in the upper right of the figure. Both galaxies have MIRI coverage constraining the NIR part of the SED.}
    \label{fig:z6_candidates}
\end{figure}

Finally, we explore the high-redshift end of our massive quiescent galaxy sample. 
Fig. \ref{fig:z6_candidates} shows SEDs and RGB images of our two $z>6$ quiescent galaxies as identified by our selection criteria. For more details on our selection criteria for the highest redshift candidates, see Appendix \ref{app.s.high-z_selection}.

The first key aspect is the recovery of 66283 which has previously been identified and published in \citet{Weibel2024qgal} and is a spectroscopically confirmed quiescent galaxy that lies at $z=7.3$. We note that our best fit \eazy\ photometric redshift is off by around 0.3, but that this is due to the difficulty of constraining a precise redshift from the Balmer break alone. Due to our wide redshift bins for the highest-z massive quiescent galaxies, this does not significantly affect our results.
Our other $z>6$ target is 30744 at $z=6.1$ which shows clear Balmer break in the photometry and has the benefit of a MIRI detection at 770$\mu $m. Our-best fit \pipes model gives it a stellar mass of $M_*=10^{9.7}M_\odot$.
This makes it a robust candidate that would be worth following up in the future with spectroscopy.

\section{Discussion}

\subsection{Observations}
In this work we have explored a large sample of massive quiescent galaxies from $z=2-7$, investigating their number densities, stellar mass functions and cosmic stellar mass density. 
Previous JWST studies have reported significantly larger number densities compared to previous works and simulations \citep[e.g.][]{Carnall2023, Valentino2023, Alberts2024, Long2024, Baker2025a}. It is complicated to compare like with like due to the number of differences between selection and SED modelling, and how these affect the number density values. 

As has been shown previously \citep[e.g.][]{Antwi-Danso2023,Baker2025a}, and comprehensively in this work, selections based on colours and sSFR give different results, especially at high-z. 
{By extending our colour selection and then using an sSFR selection, we showed that we obtain a larger population of lower-mass and higher-redshift quiescent galaxies than were found in previous works \citep[e.g.][]{McLeod2021, Weaver2023}. We also showed how galaxies that fall in an expanded region of the UVJ diagram occur at all redshifts (see Fig. \ref{fig:uvj}), so this is not simply a problem at the highest redshifts where we lose J-band coverage. This indicates that previous works based on UVJ selection were missing this population of galaxies, explaining the discrepancy between colour selected and sSFR works (see Fig. \ref{fig:smf_field}). In light of this, we highlight the importance of switching from simple colour selection to other selection criteria, such as sSFR cuts, or spectral information (where available) like the strength of the D$_n$4000 break.  }

In addition to this, the actual mass cut for selecting massive quiescent galaxies varies between works with some studies using \Mstar $> 10^{10.6}\rm M_\odot$ and others \Mstar $> 10^{10.0}\rm M_\odot$. {The specific star-formation rate selection (if used) also varies with some works using a flat cut \citep[e.g.][]{Lagos2025, Chittenden2025} or others using a threshold that evolves with redshift \citep[e.g.][]{Schreiber2018,Carnall2023, Baker2025a, Lim2025}.}
{In this work, we have tried to homogenise as much as possible by providing two different mass cuts, a \Mgnine cut and an \Mgten cut (the latter providing more literature comparisons) alongside the evolving sSFR criteria \citep{Franx2008, Carnall2018}}. 

Regarding the number densities measured, we find strong agreement with previous works such as \cite{Valentino2023} and \cite{Baker2025a}. This is likely due to the expanded quiescent galaxy selection criteria within these works.
We find a greater number of massive quiescent galaxies than \citet{Nanayakkara2025}, but fewer than \citet{Carnall2023} found in the CEERS field. 
At the high-z end, we find roughly the same number density as that reported in \citet{Weibel2024qgal}, but again caution that their reported value is based on a single source.

What we are learning from this is the importance in both probing large volumes to combat cosmic variance, but also that we need a combination of different fields which are better able to sample the underlying large-scale structure, making the number densities more robust to the effects of over- or under-densities \citep{Valentino2023,Jespersen2025, Jespersen2025b}. 

This is seen again in the SMFs. We test the effects of cosmic variance and colour selection on our results, finding field-to-field variations \citep[again reiterating the result of][ but this time for the high-z quiescent galaxy SMFs]{Valentino2023} that become increasingly pronounced at higher z. We also find significant differences based on our selection criteria, showing how just using colour selection can lead to significantly reduced SMFs on the low-mass or highest-z end (compared to an expanded colour selection followed by detailed SED modelling).

The "knee" of the fitted SMFs remains within $\rm 10^{10.3-10.5}$ regardless of redshift, suggesting that the quenching processes are operating consistently at the same mass ranges, i.e. the turnover between quenching regimes is redshift independent. 
We see tentative signs of upturns at the low-mass end for our observed SMFs, particularly in the individual fields (see Fig. \ref{fig:smf_field}, particularly COSMOS at $z=2.5-3.0$ and the combined SMF in the lower redshift bins). Whilst we caution that we may be affected by bin edge effects and low-number statistics, this is possibly a sign of environmental effects playing a role in quenching up to $z\sim 3$. Indeed, more local galaxies with stellar masses
$10^9-10^{10}\,\rm M_\odot$ would be prime targets for environmental quenching \citep{Peng2010, Hamadouche2024}. This would also explain why we see more field-to-field variation in the lower-mass end of the SMFs -- this is simply due to the varying sizes and environments of the individual fields. 

This further motivates combining multiple wide-area JWST surveys with many different filters that probe various fields to better sample the underlying large-scale structure. In addition, shallower, wide field surveys such as COSMOS-Web \citep[][]{Casey2023} will be important in understanding variations within particular fields (albeit with a reduction in available bands and depth). In addition to JWST, instruments such as NISP and VIS on EUCLID \citep{EuclidCollaboration2025, Weaver2025}, while significantly more shallow and with fewer bands, will also be vital in studying the effects of cosmic variance on the SMF. 

We integrate our SMFs in order to compute the cosmic SMD of massive quiescent galaxies and compare to other observational results \citep{Santini2021, McLeod2021, Weaver2023}. We find a sharper rise in the mass density from $z=4$ to 2 than has been seen previously \citep[][]{Weaver2023}, highlighting that we see a steep increase in the amount of mass locked up in these quiescent systems in the period up towards cosmic noon. We find that it increases by a factor of around $\times$60 from $z=4.5$ to $z=2$, such that $\rho_* \propto (1+z)^{-7.2\pm0.3}$. This increase could be due to a dramatic rise in the amount of galaxies being quenched, perhaps via stronger quenching mechanisms. Another alternative is that it may be a product of there being more massive galaxies available to quench due to the steep rise in the cosmic SFRD within this epoch \citep{Madau2014}.

In accurately determining quiescent galaxies from photometry, increased band coverage is crucial for helping remove interlopers such as Brown Dwarfs and LRDs at the highest redshifts (e.g. above $z=5$). Dedicated medium bands surveys among pre-existing fields are incredibly useful for this \citep[e.g. ][]{Williams2023,Eisenstein2023jof, Suess2024}. Greater understanding of the interloper populations will also be vital to better understand and remove them.

\subsection{Simulations}
Predictions differ wildly between simulations. With a mass cut of \Mgten we find poor agreement with most cosmological hydrodynamical simulations and SAMs. \flamingo \citep{Schaye2023} and \tng \citep{Pillepich2018,Springel2018} perform poorly, with neither of these works having close to enough massive quiescent galaxies, failing to reproduce the observed number densities above $z=3$. \tng does however manage to reproduce the observed number densities at $z=2-3$. Interestingly, \flamingo, \simba\citep{Dave2019}, and \eagle \citep{Schaye2015, Crain2015} all fail to produce the number densities at $z=2-3$, despite this range being the best-constrained by observations. \galform \citep{Lacey2016} and \tng struggle at higher-z, i.e. above $z=3.5$. 
\shark \citep{Lagos2018, Lagos2024} performs better, generally matching the number densities up to $z=4$, but deviates from our observations at higher redshifts. {\magneticum shows that it is possible to produce many quenched galaxies, however, it massively overproduces them, severely overpredicting the number densities within the redshift range $z=2-5$ which are the most tightly constrained by observations.}

We also explored including errors in the simulations to better mimic observations, finding that for most simulations, while this changes the number densities, it cannot account for the discrepancies seen (although this pushes \shark to accurately predict the number densities up to $z=6$).
This highlights the difficulty in obtaining accurate number densities at all redshifts with simulations.

A more in-depth question is whether the simulations can reproduce the observed mass distribution of these massive quiescent galaxies and this is tested by the stellar mass function.
We find that none of the simulations appears to reproduce the quiescent galaxy stellar mass function as observed in this work (or in \citealt{Weaver2023}) at any redshift (although \tng is close at $z=2-3$). Predictions are again inconsistent with some simulations obtaining enough high-mass quiescent galaxies, but failing at the low-mass end, whilst others can reproduce the low-mass, but fail at the high-mass end. These discrepancies still hold after adding gaussian-distributed errors on simulated quantities (this just has the effect of broadening the simulated SMFs.)

We also see that several simulations fail to fully reproduce the SMD of quiescent galaxies across cosmic time. Interestingly, most remain within 2\sig of the observations (with the exception of \eagle and \magneticum), and the dramatic differences seen in the SMF are mostly washed out, with most simulations producing SMDs within 0.3~dex of each other. \eagle dramatically under-predicts the SMD {at all redshifts,} whilst \magneticum significantly overpredicts it.
\tng fails at $z>3.5$ due to a lack of quiescent galaxies. Despite obtaining the closest number densities, \shark overpredicts the SMD at $z>3$ due to its significantly larger population of lower mass quiescent galaxies. \simba and \galform appear mostly able to reproduce the observed SMD (despite struggling with the number densities and SMFs).

As has previously been reported in \citet{Lagos2025} for $z=3$, these findings for the simulations are likely due to the varying feedback prescriptions with them, {particularly the onset and strength of AGN feedback.} Simulations where it only kicks in above a certain mass range (e.g. \tng) struggle to reproduce any of the less massive quiescent galaxies. Meanwhile, other simulations such as \eagle well reproduce the low-mass end of the SMFs but cannot produce high-mass quiescent galaxies, showing that their high-mass quenching mechanisms need stronger or more efficient feedback. We see from the SMFs that the lower mass quenching mechanisms appear to be much too strong within the SAMs \shark and \galform. \simba fails to reproduce the shape of the SMF at either low or high-mass ends, suggesting both require adjustments.

Resolution effects and box-sizes are also important. As has been shown previously in \citet{Baker2025a}, simulation based number densities are also subject to cosmic variance because of their inherent small box sizes. Whilst large volume simulations exist they often struggle to match the required small-scale physics to produce enough massive quiescent galaxies \citep[as seen for \flamingo in][]{Baker2025a}. This suggests that large-volume simulations are crucial to minimise the effects of cosmic variance even within the simulations.

On the other hand, quenching both massive and low-mass galaxies is a multi-scale problem \citep{Man2018, DeLucia2025} with many of the crucial feedback processes taking place at much smaller scales. This motivates further research into the quenching mechanisms within high resolution simulations.
However, this proves to be a big challenge for simulations as they can either aim for a large box (sacrificing resolution and resorting to sub-grid physics) or they can resolve the small-scale physics but only for a handful of galaxies which is too small for number densities or SMFs \citep[for a review of the problem see][]{Vogelsberger2020}. Small-scale simulations also struggle to find massive quiescent galaxies at high z due to their relative rarity within the overall galaxy population. Hence, cosmological simulations also need more complicated feedback prescriptions in order to reproduce observations.

Overall, future wide and deep JWST surveys (combined with legacy data) and the next generation of cosmological hydrodynamical simulations and SAMs will start to address these issues.

\subsection{Caveats and limitations}

{Within any work of this kind there are necessarily caveats and limitations. These correspond to both the observational part of the analysis and the simulations. We will focus on the observational part here and refer readers to \citet{Lagos2025} for the simulations. One of our (and many other works) primary caveats is the SED modelling. This takes two forms within our work - the initial \eazy\ template fitting run and then the more detailed Bayesian \pipes fitting. The usual caveats corresponding to SED modelling apply \citep[e.g. ][]{Conroy2013}. We note however, that as we are exploring massive quiescent galaxies aspects such as outshining \citep[e.g.][]{Sawicki1998} should be insignificant in comparison to other works involving star-forming galaxies. The usual caveats about systematic effects apply however, such as star-formation histories \citep{Leja2019, Wang2023}, metallicities, IMFs \citep[e.g.][]{vanDokkum2024} and abundance ratios \citep{Beverage2025}. }

{Uncertainties on the photo-zs for these type of massive quiescent galaxies are another source of uncertainty but have been shown to be sub-dominant in \cite{Ito2025b}, where they cover the lower redshift range of our study, although photo-z uncertainty is more significant for our highest-z candidates. We also note that we fix the \pipes redshift to that of the \eazy\ photo-z as opposed to allowing it to vary \citep[e.g.][]{Weibel2024}, as it has been shown to be more accurate at recovering spectroscopic redshifts \citep{Weibel2024,Ito2025b}. This may have the effect of ruling out potential areas of the Bayesian posterior space in \pipes, but, due to the improved accuracy of the \eazy\ photo-zs, it should primarily rule out less feasible solutions.}

{Another key aspect is the role of AGN within these galaxies. Recent spectroscopic JWST works have found a number of high-z massive quiescent galaxies show signs of current AGN activity \citep[][]{Bugiani2025, D'Eugenio2024, Carnall2023Nature,Baker2025a} which alter their SEDs compared to non-AGN host galaxies. We do not include AGN models within our \pipes SED modelling. This could increase or reduce the number of quiescent galaxies within our sample. Firstly, there is the possibility for interlopers due to extreme emission lines at certain redshifts which could mimic a Balmer break. Specifically in our case this is an issue at the highest redshifts due to the prevalence of LRDs \citep{Matthee2024} which we do our best to remove via visual inspection. Secondly, solutions involving emissions lines within \pipes will attribute those emission lines to star-formation rather than AGN activity, potentially misclassifying AGN activity in quiescent galaxies as SF, leading to an underestimate. The first of these should be minor for a population study of this size, at least up to $z\sim$5, but is likely to be a bigger problem at high-z. This will require larger spectroscopic surveys to investigate in depth. The second should be lessened by our extended UVJ criteria in the initial \eazy\ selection.}

We also note that we are exploring the integrated properties of these galaxies and on spatially resolved scales individual properties can vary due to differing morphological components, stellar populations, dust properties and other evolutionary processes \citep[e.g.][]{Tacchella2015,D'Eugenio2024,Mosleh2025,Baker2025}.

To summarise, we have the same uncertainties and caveats as expected for studies involving integrated SED modelling of quiescent galaxies. We do not expect these to be significant within the bulk of the analysis, which is restricted primarily to $z=2-5$ where we have excellent number statistics. However, above redshift 5 where we are restricted to a handful of candidates, these are likely to be a cause of more uncertainty. 

\section{Conclusions}
We have assembled a robust sample of over 700 massive quiescent galaxies from $z=2-7$ from a combination of some of the deepest JWST multi-band surveys with an overall area 5$\times$ larger than existing JWST studies. 
We computed quiescent galaxy number densities at redshifts $z=2-7$ and explored stellar mass functions and the cosmic stellar mass density from $z=2-5$. We compare with a range of simulations and SAMs with matching selection criteria to provide an accurate comparison to the state-of-the-art galaxy evolutionary modelling.
Our key findings are as follows.

\begin{enumerate}
    \item We confirm an overabundance of massive quiescent galaxies relative to pre-JWST works and models at high-z. 
    \item We find significant numbers of observed massive quiescent galaxies that would be missed by the traditional UVJ colour selection criteria, further motivating cuts based on sSFR.
    
    \item {We find that the cosmological hydrodynamical simulations we explore cannot accurately predict the observed number density of massive quiescent galaxies in the range $z=2-7$.} 
    
    \item Simulations cannot reproduce the observed quiescent galaxy SMF at any redshift from $z=2-5$. All fail to match the shape and normalisation of the observed SMF. Some models overpredict lower-mass quiescent galaxies (i.e. \shark, \galform) and fail at the high-mass end, whilst others such as \tng match the high-mass end but fail to quench any lower-mass galaxies. This is connected to the various feedback prescriptions in the simulations.
    
    \item Quiescent galaxy stellar mass functions can vary significantly with differing selection criteria, and we uncover more quiescent galaxies at the lower-mass end than previous studies. This is due to our extended colour selection and our sSFR cut being better able to uncover the lower-mass quiescent galaxy population. 
    \item We show the impact of field-to-field variation in SMFs due to cosmic variance and demonstrate how it significantly increases at higher redshift. We also show significant variations when different colour selection criteria are used compared to a sSFR cut.
    \item We find that the stellar mass density of quiescent galaxies evolves more steeply between $z=5$ and $z=2$ than has been found in previous work, with $\rho_*\propto (1+z)^{-7.2\pm0.3}$, indicating a significant increase in the importance of galaxy quenching within these epochs.
    \item As part of the sample, we report another robust high-z ($z>6$) quiescent galaxy candidate.
    
\end{enumerate}

Our best models for galaxy evolution, as encapsulated by current cutting-edge cosmological simulations, are unable to accurately reproduce the high-z quiescent galaxy population as observed. This presents a wonderful opportunity to improve our knowledge of high-z galaxy evolutionary theory with the next generation of cosmological simulations.


\begin{acknowledgements}

WMB thanks Lucas Kimmig and Rhea-Silvia Remus for providing data from Magneticum Pathfinder for the comparisons, and also for useful discussions.
      
In addition, WMB would like to thank Francesco D'Eugenio and Marko Shuntov for helpful and enlightening discussions. 

WMB would like to acknowledge support from DARK via the DARK Fellowship. This work was supported by a research grant (VIL54489) from VILLUM FONDEN.
Some of the data products presented herein were retrieved from the Dawn JWST Archive (DJA). DJA is an initiative of the Cosmic Dawn Center, which is funded by the Danish National Research Foundation under grant DNRF140. FV and KI acknowledge support from the Independent Research Fund Denmark (DFF) under grant 3120-00043B. This study was supported by JSPS KAKENHI Grant Number JP23K13141.
AS is supported by a Villum Experiment grant (VIL69896) and research grant (VIL54489) from VILLUM FONDEN.

We acknowledge use of \textsc{astropy} \citep{AstropyCollaboration2013}, \textsc{fsps} \citep{Conroy2009, Conroy2010}, \eazy\ \citep{Brammer2008}, \textsc{grizli} \citep{Brammer2023}, \pipes \citep{Carnall2018, Carnall2019}, \textsc{numpy} \citep{harris2020array}, \textsc{dynesty} \citep{Speagle2020}, 
\textsc{matplotlib} \citep{Hunter:2007}, and \textsc{topcat} \citep{Taylor2005}.

This work is based [in part] on observations made with the NASA/ESA/CSA James Webb Space Telescope. The data were obtained from the Mikulski Archive for Space Telescopes at the Space Telescope Science Institute, which is operated by the Association of Universities for Research in Astronomy, Inc., under NASA contract NAS 5-03127 for JWST. These observations are associated with programs \#1345, \#1837, \#1180, \#1181, \#1210, \#1287, and \#1215.

\end{acknowledgements}

\bibliographystyle{aa}
\bibliography{refs}

\begin{appendix}

\section{\pipes setup}

\label{app.s.pipes_setup}

Here we explain in more depth our \pipes setup.
We use the updated \citet{Bruzual2003} stellar population models \citep{Chevallard2016} alongside the MILES library of stellar spectra \citep{Sanchez-Blazquez2006,Falcon-Barroso2011} and the updated stellar tracks from \citet{Bressan2012, Marigo2013}. Nebular components, i.e. line and continuum emission is incorporated following the procedure in \citet{Byler2017}. We enable the ionisation parameter U to vary between -4.5 and -0.5 in log space. We use a \citet{Calzetti2000} dust attenuation prescription with the dust extinction in the V band allowed to vary between 0-8 magnitudes. 

For our star-formation history we use a parametric double powerlaw prescription \citep{Carnall2018, Carnall2019}. There is much debate about how best to model high-z star-formation histories \citep[see][for more of a review of the various methods]{Carnall2019parametric, Leja2019}. We chose a double powerlaw model due to the increased speed in model fitting and since we are primarily fitting broad band photometry we are not particularly sensitive to the exact details of the star-formation history. It should be noted that increases in stellar masses are found when using a more flexible "continuity" SFH \citep{Leja2019stellarmasses}, however it remains unclear whether the "continuity" prior used in the fitting is really physically motivated at high-z \citep[see][]{Wang2023,Turner2025, Mosleh2025}. 
In light of this we stick with the more simple prescription. We use a uniform prior between 0.01 and 1000 for the indices $\alpha$ and $\beta$ in the double powerlaw SFH. We leave log stellar metallicity free to vary between -2.5 and 0.19 (i.e. $\log(Z/Z_\odot)=(-2.5,0.19)$), whilst noting that we cannot constrain this parameter from photometry. During our initial run we left the redshift free to vary and then compared the results for 18 spectroscopically confirmed galaxies from \citet{Baker2025a} (see Appendix \ref{app.s.ComparisonSpectra}). We found that the \eazy\ redshifts were more accurate than the \pipes redshifts so we fix the \pipes redshifts to that of the \eazy\ values.

\section{Completeness}
\label{app.s.completeness}

\begin{figure}
    \centering
    \includegraphics[width=1\linewidth]{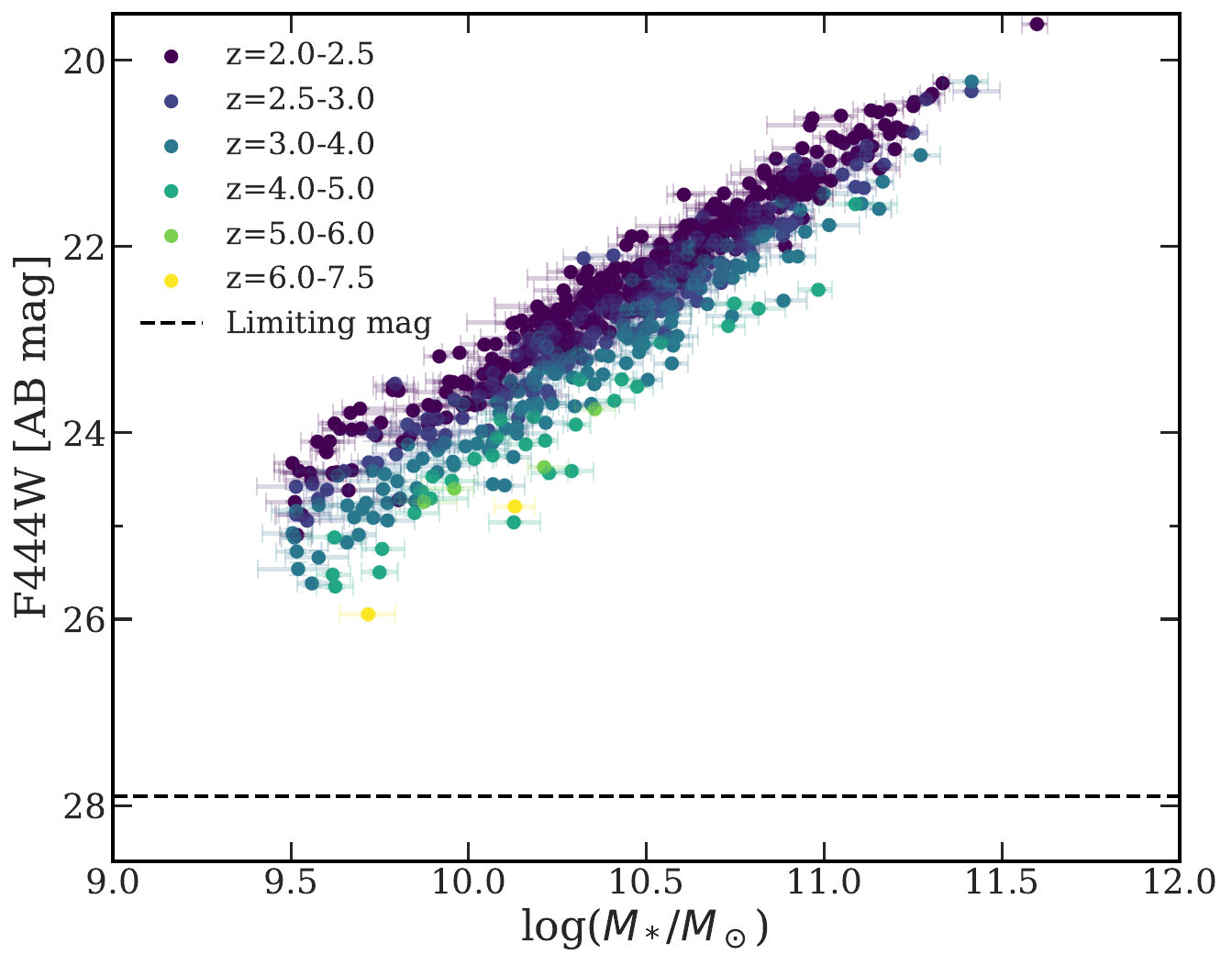}
    \caption{Observed F444W magnitude versus stellar mass as obtained via \pipes SED modelling colour coded in bins of redshift for our sample. It shows that we are not missing any quiescent galaxies within our analysis due to being magnitude limited. }
    \label{fig:completeness}
\end{figure}

Fig. \ref{fig:completeness} shows the F444W magnitude versus stellar mass (obtained via fitting with \pipes), colour-coded by bins of redshift for our entire sample. The key takeaway is that due to our mass cut of $M_*>10^{9.3}M_\odot$ for the SMFs and $M_*>10^{9.5}M_\odot$ for the number densities we are able to see any massive quiescent galaxy within the survey area. The observed F444W magnitude versus stellar mass figure shows that, due to our mass cut, we are clearly complete and are not missing sources due to being magnitude limited. 
We also see the standard expected evolution with redshift, with the higher redshift sources appearing fainter.

\section{Stellar mass function fitting}
\label{app.s.smf_fitting}
\begin{figure*}
    \centering
    \includegraphics[width=0.49\linewidth]{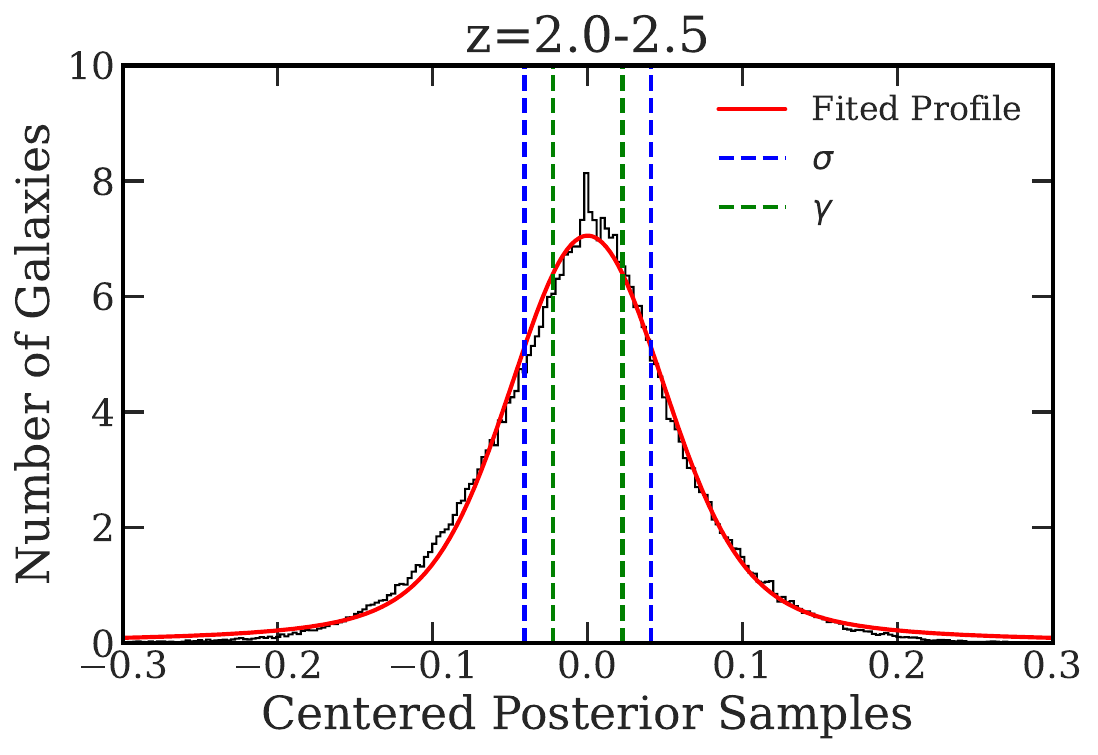}
    \includegraphics[width=0.49\linewidth]{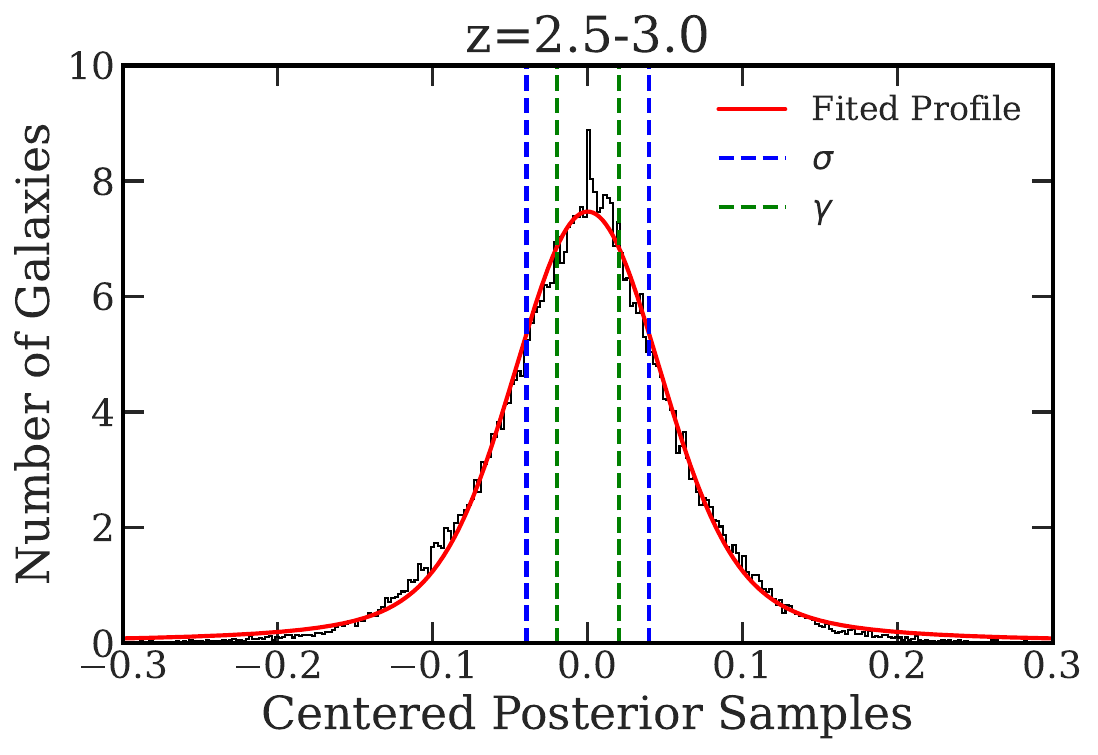}
    \includegraphics[width=0.49\linewidth]{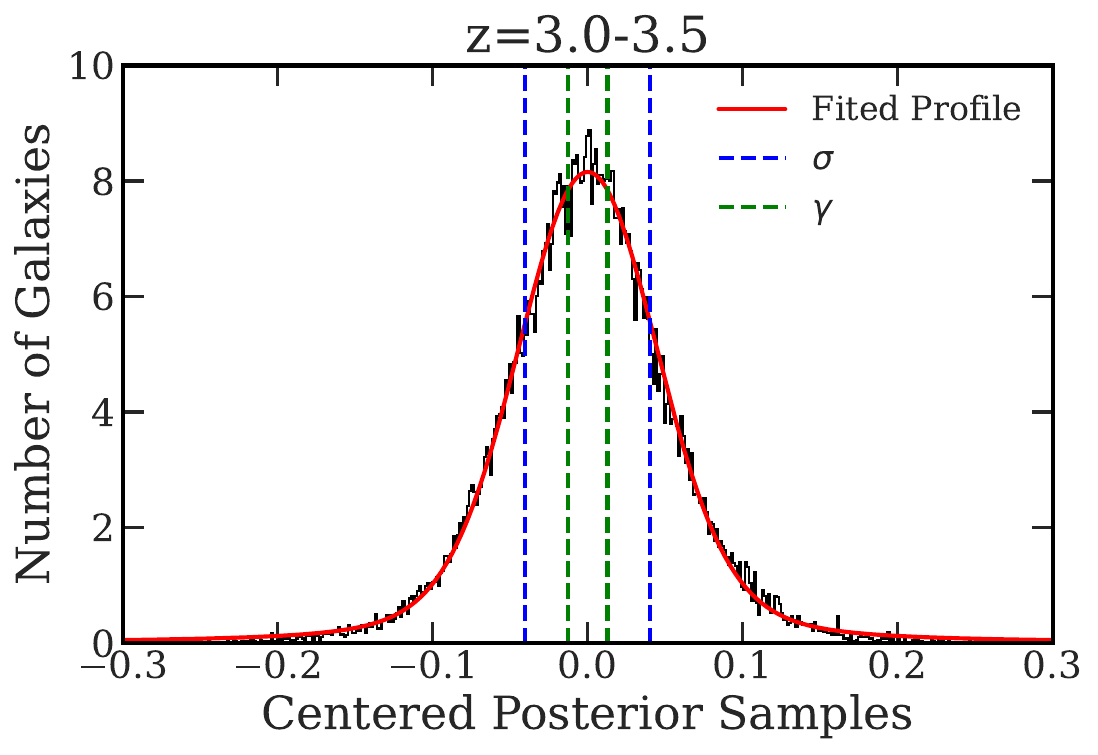}
    \includegraphics[width=0.49\linewidth]{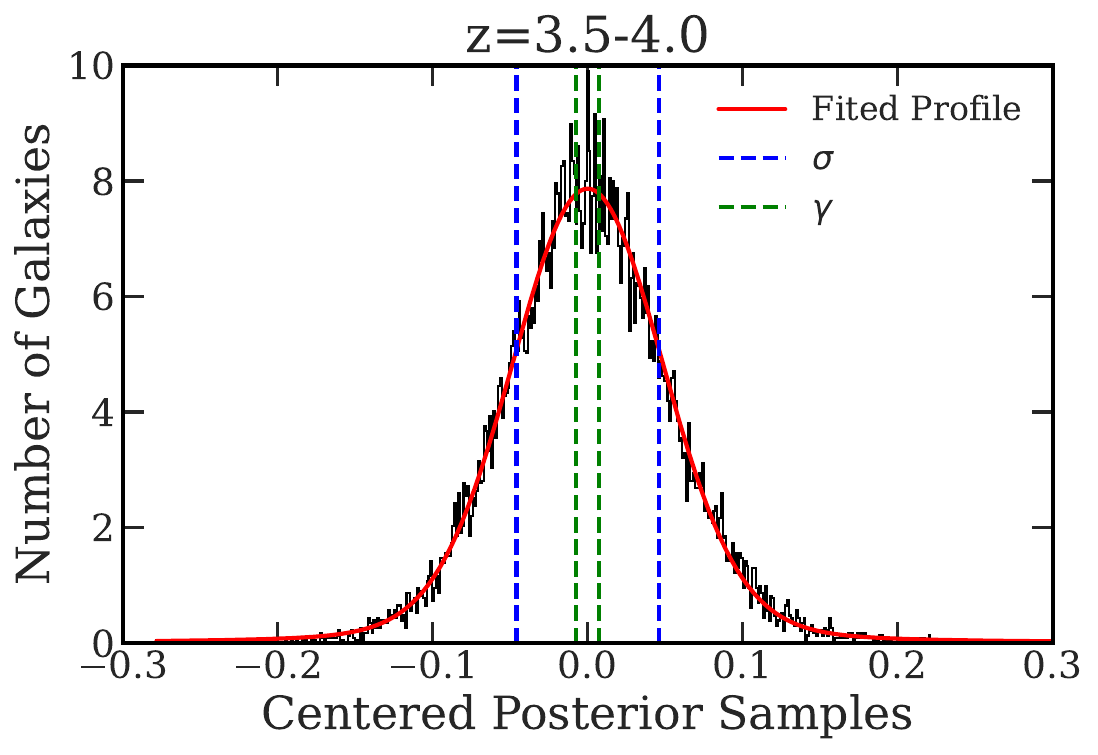}
    \includegraphics[width=0.49\linewidth]{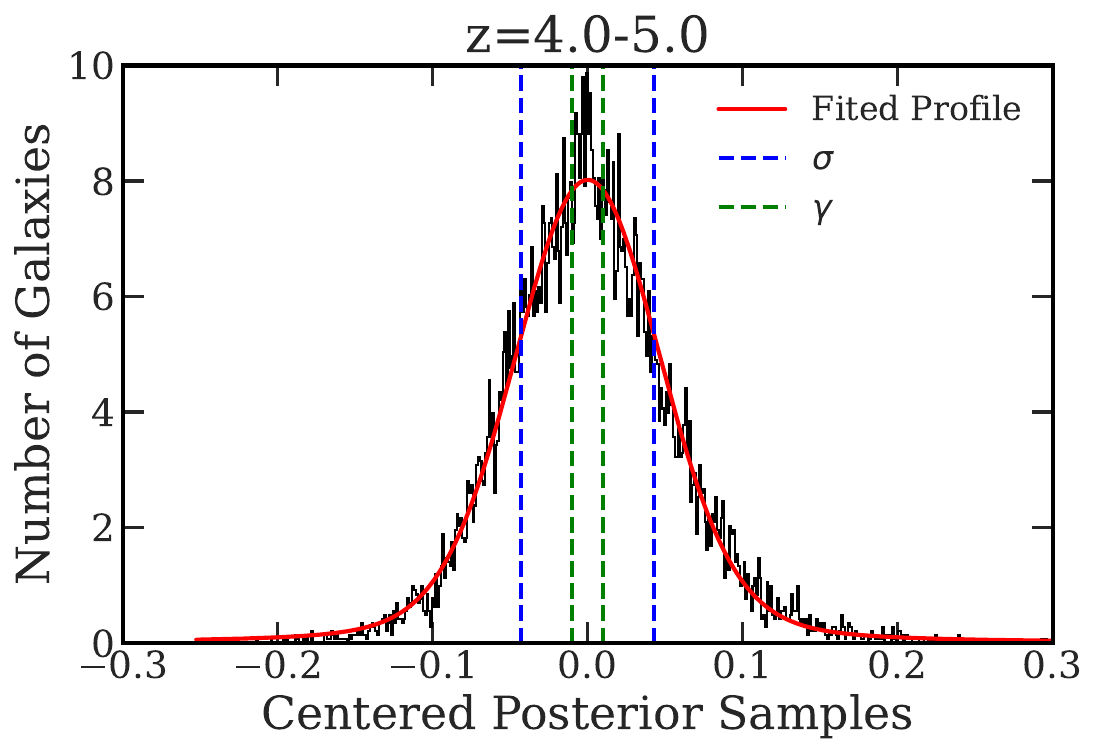}
    \caption{Density kernels produced by stacking the normalised log stellar masses of all galaxies within each redshift bin. These are fit with a Voigt profile which is then convolved with the SMF for that redshift in order to account for \citet{Eddington1913} bias.}
    \label{fig:kernels}
\end{figure*}

The first aspect to note is that uncertainties on individual quantities of course end up affecting the SMF itself. For example, errors on redshifts and stellar masses resulting from SED modelling can lead galaxies to be in the wrong bins, which can be hard to diagnose since SED codes generally produce overconfident posteriors due to model misspecification \cite{Jespersen2025_optical-IR}. Therefore we consider typical uncertainties/biases derived from other studies. Generally photometric redshifts are measured accurately due to the number of bands and flexibility of templates, but there are offsets of 0.1-0.2 dex \citep[][and see Appendix \ref{app.s.ComparisonSpectra}]{Baker2025a}. Stellar masses can have significant systematic offsets of up to 0.3 dex \citep{Conroy2013}.

In the actual measurements of the SMF, there are three common forms of error \citep[e.g.][]{Adams2021, Weaver2023, Shuntov2025}.
The first of these is completeness. For a photometric study such as this one this takes the form of the well-known Malmquist bias -- we are limited in magnitude as to the faintest object we can see. In our case we have a mass cut of greater than \Mstar>$10^{9.3}M_\odot$, and as can be seen in Appendix Fig \ref{fig:completeness} our faintest object is far above our limiting magnitude. So for our purposes our sample is mass complete above \Mstar>$10^{9.3}M_\odot$

The second issue is cosmic variance. With deep extragalactic surveys we can only probe a small fraction of the sky hence our volume is necessarily limited. Whilst the Universe is likely homogeneous on large enough scales, we cannot probe those exact scales. We help minimise the effect of cosmic variance by using a large survey area (our area is approximately 6x the size of other comparable studies), but also by using fields in various different regions of the sky \citep[as in][]{Valentino2023}. In computing our observational errors we use a prescription for fractional cosmic variance for each redshift and mass bin based on the approach of \citet{Jespersen2025}. 

The final source is the so called "Eddington Bias" \citep{Eddington1913}. This occurs due to measurement errors in quantities. For our work this almost entirely corresponds to the stellar mass of the galaxies. Due to the impossibility of knowing their masses exactly, galaxies can scatter to higher or lower masses than their "true" values. If the masses were equally distributed this would result in the same number of galaxies scattering to higher stellar masses as the number scattering to lower stellar masses. However, as can be seen in most papers exploring stellar mass functions \citep[e.g.][]{Davidzon2017, Adams2021, Weaver2023, Shuntov2025} there are more galaxies at certain masses than at others. This means that this scatter has a preferred direction, meaning this result must be accounted for when fitting a stellar mass function. This can be taken into account in multiple ways. The approach we will use is to take into account the posterior samples for each galaxy fit from \pipes. We then fit a Voigt profile (a convolution of a Gaussian with a Lorentzian profile) to this normalised distribution to obtain a measure of the scatter. 
This Voigt profile is then convolved with the stellar mass function and the convolution is fit to the data. 
It is also possible to create kernels based upon the posterior distributions \citep[e.g.][]{Shuntov2025}, but we find that the Voigt profile fits our posterior distributions well (see Appendix Fig. \ref{fig:kernels}). An alternative approach is to use a likelihood based methodology when fitting \citep{Leja2019stellarmasses}, but we do not use this approach in our work due to our smaller sample size.

\section{\eazy\ fitting and posterior for high-z target}

\begin{figure*}
    \centering
    \includegraphics[width=0.9\linewidth]{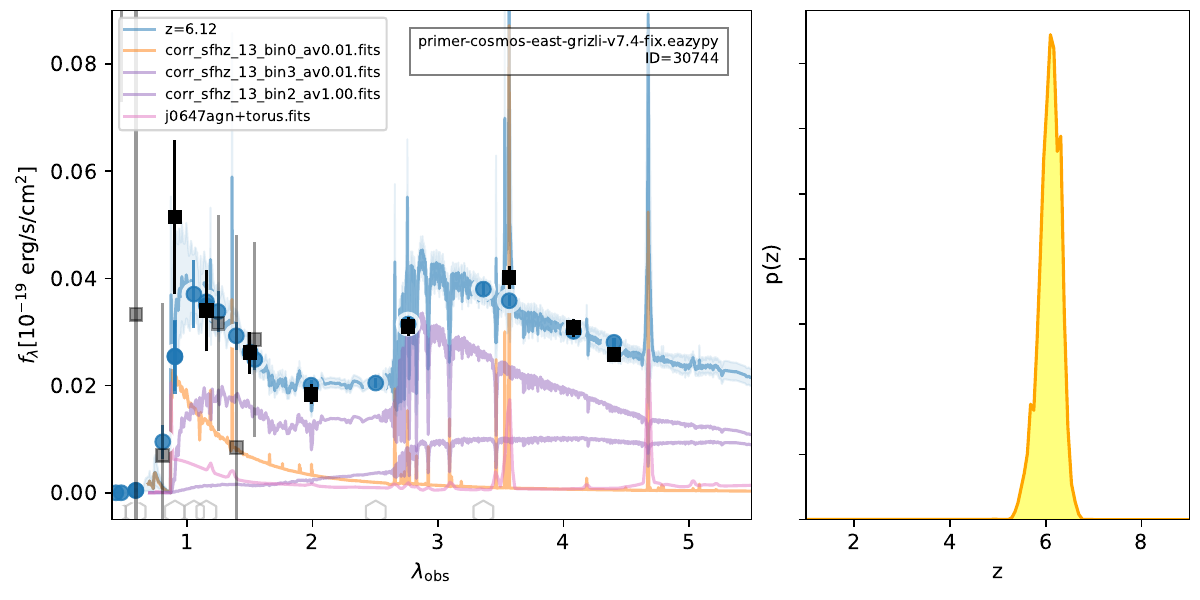}
    \includegraphics[width=0.9\linewidth]{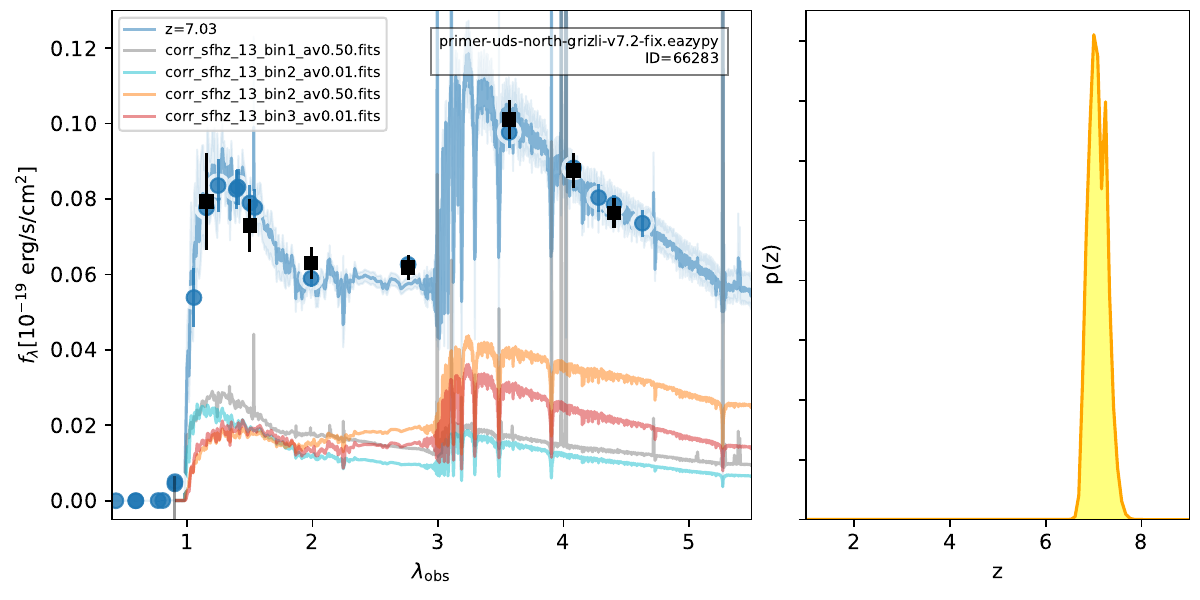}
    
    \caption{Left: Best-fit \eazy\ templates to 30744 (upper) and 66283 (lower). We note that both have MIRI constraints that are not shown here. Right: the \eazy\ redshift posterior distribution for the sources. }
    \label{fig:eazy_highz}
\end{figure*}
\label{app.s.high-z_selection}
In this section, we explore in more depth our selection for high-redshift candidates.
This is the most challenging in terms of rooting out possible contaminants due to the combined contamination of both Brown Dwarfs and LRDs. As mentioned earlier, Brown Dwarfs are removed by a colour cut following the criteria of \citet{Langeroodi2023} and \citet{Kokorev2024} with any galaxy whose colour falls within the tracks of the modelled Brown Dwarfs being removed. The effect of this cut is to remove galaxies at $z>5$ with both F150W-F200W$<0.2$ mag and  F200W-F277W$>1$ mag. However, we acknowledge with this method it is possible that Brown Dwarfs still remain due to uncertainties in the blue bands of the photometry, although we have done our best to mitigate this with the visual inspection. 

LRD contamination remains more challenging as they could be either AGN or quiescent galaxies and the populations exact nature is unconfirmed at this time. {We want to minimise contamination from sources with significant non-stellar flux.}
At these redshifts almost all these sources are point-like so in some ways the difference between a LRD and a quiescent galaxy is a tricky question, especially as there is only one spectroscpically confirmed quiescent galaxy at $z>5$ and it is only marginally resolved \citep{Weibel2024qgal}. Our approach is to remove galaxies based on visual inspection of their SEDs and best fit \pipes models. We do not use any AGN templates in our \pipes SED modelling \citep[although note that this is possible, e.g.][]{Carnall2023Nature}, hence we remove galaxies that appear poorly fit, likely due to an AGN contribution to the continuum or other contamination effect. We also remove galaxies with a clear "V" shaped SED.

In Fig. \ref{fig:eazy_highz} we show the best-fit \eazy\ model and the posterior for our high-z candidate 30744 which is found within the Primer Cosmos field. We also show the same for 66283 which is the only spectroscopically confirmed high-z quiescent galaxy \citep{Weibel2024qgal}.
We see that the posterior of our candidate is fully consistent with a redshift of around 6. We again note that this candidate has a MIRI detection which is not shown in the best-fit \eazy\ model.

\section{Comparison to spectra}
\label{app.s.ComparisonSpectra}
\begin{figure*}
    \centering
    \includegraphics[width=0.85\linewidth]{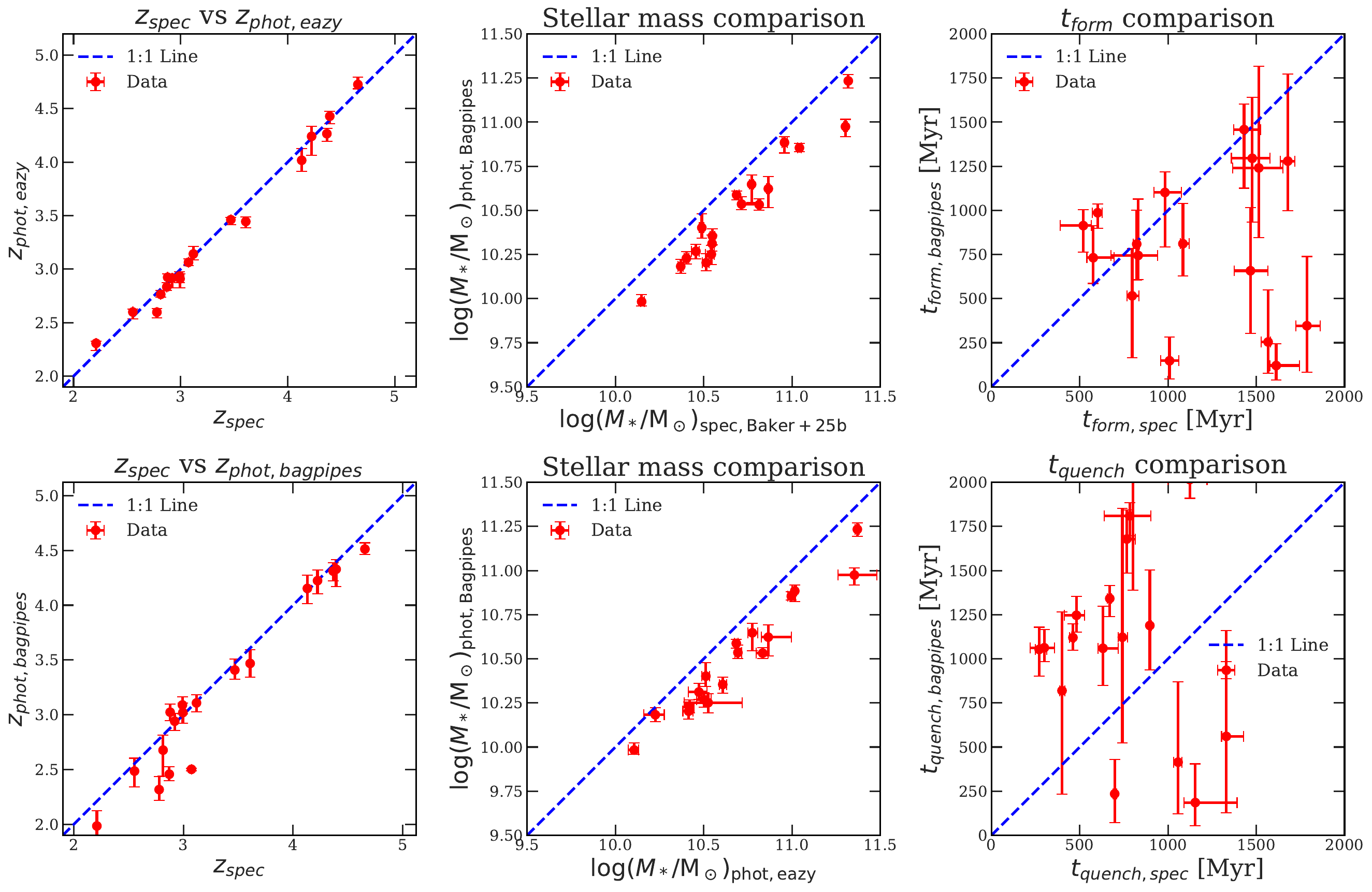}
    \caption{Comparison between quantities for a sample of 18 massive quiescent galaxies from \citet{Baker2025a} compared with our work. Upper left: comparison of our photometric \eazy\ redshifts versus the spectroscopic redshifts. Lower left: comparison of photometric redshifts from \eazy\ and those from \pipes. Upper middle: comparison between the stellar masses obtained via photometric fitting with bagpipes and those from combined spectrophotometric fitting. Lower middle: comparison between the stellar masses obtained from photometry with \pipes vs \eazy. 
    Upper right: comparison of the formation times ($t_{50}$) and lower right a comparison of the quenching times ($t_{90}$).}
    \label{fig:spectroscopic comparison}
\end{figure*}

We can compare our SED fit quantities to other spectroscopic samples in order to see how well we are able to recover quantities from photometry alone. We compare to the 18 massive quiescent galaxies from \citet{Baker2025a} who used both NIRSpec PRISM spectra and photometry in their SED modelling. 
The results of this is shown in Fig. \ref{fig:spectroscopic comparison}. The upper left panel shows that our \eazy\ photometric redshifts are accurate for these galaxies with no significant offsets. {The lower left panel shows the results of fits where the redshift was left free in \pipes, compared to the spectroscopic redshift. We see increased scatter between the spectroscopic redshifts and the photometric redshifts from \pipes (lower left) compared to those from \eazy\ (upper left).}
Our stellar masses are slightly lower (upper middle panel), but this is also likely due to the different SFHs between the \citet{Baker2025a} work and our own \citep[for more details between offsets in SFHs see][]{Leja2019}, alongside the differences between PSF matched and non-PSF matched photometry. 
However, we see that we obtain significant differences in the formation and quenching times.{With the assumption that the spectroscopic formation and quenching times are closer to the "true" value, this suggests that with photometry alone we
do not want to place undue importance on the formation and quenching time values. However, a much more detailed and comprehensive study is required to make significant conclusions on the accuracy of these values measured from photometry alone. This is because our comparison is still affected by the usual SED modelling assumptions, including different fitting codes and their stellar isocrones and stellar libraries, SFH models, and various other model parameters. For this work therefore, we shall simply accept that we do not have enough precision to explore formation and quenching times for our sample.}  This motivates further spectroscopic and photometric exploration of large samples of massive quiescent galaxies in order to understand their SFHs on a population level.

\end{appendix}

\end{document}